\documentclass[conference]{IEEEtran}
\usepackage{cite}
\usepackage{amsmath,amssymb,amsfonts}
\usepackage{algorithmic}
\usepackage{graphicx}
\usepackage{textcomp}
\usepackage{xcolor}
\usepackage{soul}
\usepackage{fancyhdr}
\usepackage[hyphens]{url}

\def\BibTeX{{\rm B\kern-.05em{\sc i\kern-.025em b}\kern-.08em
    T\kern-.1667em\lower.7ex\hbox{E}\kern-.125emX}}

\fancypagestyle{firstpage}{
  \fancyhf{}

  \fancyhead[C]{\normalsize{\copyright IBM Corporation
      \textbf{-- DRAFT under review} }} 
  \fancyhead[C]{} 
  \fancyhead[C]{\normalsize{\copyright IBM Corporation}} 
  \fancyfoot[C]{\thepage}
}

\usepackage{hyperref}

\usepackage{listings}

\newcommand{\acc}{{\sf ACC}}
\newcommand{\vsr}{{\sf VSR}}
\newcommand{\mme}{{\sf MME}}
\newcommand{\instr}[1]{{\emph{\bf \sf #1}}}
\newcommand{\code}[1]{{\tt #1}}
\newcommand{\dgemm}{{\sf DGEMM}}
\newcommand{\builtin}{built-in}
\newcommand{\builtins}{built-ins}
\newcommand{\Builtins}{Built-ins}

\newcommand{\eg}{{\emph e.g.}}

\newcommand{\micro}[1]{{\sf #1}}
\newcommand{\commentout}[1]{{}}

\newif\ifcomments
\commentstrue 

\renewcommand{\hl}[1]{{#1}}

\begin{document}

\title{\vspace*{-0.8\baselineskip} A matrix math facility for Power ISA\texttrademark\ processors}

\author{
    \begin{minipage}{0.99\textwidth}
	\begin{center}
	    Jos\'{e} E. Moreira,
	    Kit Barton,
	    Steven Battle,
	    Peter Bergner, 
	    Ramon Bertran,
	    Puneeth Bhat,
 	    Pedro Caldeira,
	    David Edelsohn,
	    Gordon Fossum,
	    Brad Frey,
	    Nemanja Ivanovic,
 	    Chip Kerchner,
	    Vincent Lim,
	    Shakti Kapoor,
 	    Tulio Machado Filho,
	    Silvia Melitta Mueller,
	    Brett Olsson,
	    Satish Sadasivam,
	    Baptiste Saleil,
	    Bill Schmidt,
	    Rajalakshmi Srinivasaraghavan,
	    Shricharan Srivatsan,
	    Brian Thompto,
	    Andreas Wagner,
	    Nelson Wu
	\end{center}
    \end{minipage}
    \\
	International Business Machines Corporation, Armonk, NY, USA
    ({\tt jmoreira@us.ibm.com})
}
 

\maketitle
\thispagestyle{firstpage}
\pagestyle{plain}

\vspace*{-2.0\baselineskip}

\begin{abstract}

	Power ISA\texttrademark\ Version 3.1 has introduced a new
	family of matrix math instructions, collectively known as the
	Matrix-Multiply Assist (MMA) facility.	The instructions in
	this facility implement numerical linear algebra operations on
	small matrices and are meant to accelerate computation-intensive
	kernels, such as matrix multiplication, convolution and discrete
	Fourier transform.  These instructions have led to a power-
	and area-efficient implementation of a high throughput math
	engine in the future POWER10 processor. Performance per core is 4
	times better, at constant frequency, than the previous generation
	POWER9 processor.  We also advocate the use of compiler \builtins\
	as the preferred way of leveraging these instructions, which we
	illustrate through case studies covering matrix multiplication
	and convolution.

\end{abstract}

\section{Introduction}
\label{Sec:Introduction}

\hl{
    The IBM POWER10 processor}~\cite{9352481,9220618} \hl{is the compute engine for the next generation of
Power Systems and successor to the current POWER9}~\cite{power9-ieeemicro2017,power9-core-jrd2018} \hl{processor.  As such,
it has to offer superior performance on applications of interest to
Power Systems users. These include traditional scientific and engineering
applications and, even more important from a market perspective, business
analytics applications.
}

\hl{
Business analytics applications often rely on numerical linear algebra
computations. Important algorithms for business analytics include
classical machine learning (ML), such as liner regression, principal
component analysis, and collaborative filtering. More recently, there
has been growing (but still lagging behind ML) interest in deep learning
(DL) algorithms, including convolutional neural networks. Both classes of
algorithms are heavy users of numerical linear algebra, with ML tending
to favor the more traditional, scientific computing-like, IEEE single
(32-bit) and double (64-bit) precision arithmetic, whereas DL favors
a mix of single and reduced (16-bit floating-point, 8-bit integer)
precision arithmetic.
}

\hl{
Business analytics applications can be broadly classified as either
operating off-line (data-at-rest, as in a repository) or in-line
(data-in-flight, as in during a transaction). Operations on data-at-rest
tend to be of a larger scale and do well with attached accelerators such
as GPUs and TPUs. Operations on data-in-flight tend to be of a smaller
scale individually (although the total amount of data and computation
are often very large) and benefit from better execution speed in
the processing core performing the transaction. A system processing
data-in-flight is likely to be evaluating multiple distinct models at
once, one (and sometimes multiple) for each transaction. Agility and
flexibility of switching models, while performing well, are important.
}

\hl{
In support of these usage scenarios, the POWER10 processor had to offer
world-class performance on a spectrum of numerical linear algebra kernels,
covering both conventional as well as reduced precision arithmetic. It had
to process a large number of independent business analytics calculations,
as well as some very large scientific and technical computations. The
processor was being designed with four vector pipelines per core. When
combined with a high-bandwidth memory system, from cache to main memory,
it had sufficient throughput for BLAS1- and BLAS2-class computations.
}

\hl{
The POWER10 processor still needed additional performance on BLAS3-class
computations. The timeline of the project, and the realities of Silicon
technology, imposed various constraints on the design. Expanding the ISA
with additional, fully architected register space was not an option. The
solution would have to make do with $64 \times 128$-bit vector-scalar
registers and 128-bit wide vector instructions. Developing new compilation
technology was also out of the question. There was no time to upgrade
the operating systems to increase architected state.
}

\hl{
The solution adopted is a new facility introduced in Power
ISA}\texttrademark\ \hl{Version 3.1: The VSX Matrix-Multiply Assist
(MMA) instructions}~\cite{PowerISA3.1}. \hl{These instructions directly
implement rank-$k$ update operations on small matrices and vectors and
can be used to speed up the execution of critical dense linear algebra
kernels.  In a rank-$k$ update operation, an output matrix is updated with
the product of two input vectors or matrices.
}

\hl{
The MMA instructions use the 128-bit vector-scalar registers for input
and a new set of registers called accumulators for the output. Each
accumulator has 512 bits and can hold a $4 \times 4$ matrix of
$32$-bit elements (or $4 \times 2$ matrix of $64$-bit elements).
In the POWER10 implementation of MMA, the accumulators are stored in
the functional unit that performs the rank-$k$ update operations, which
had the benefit of significantly reducing switching power.  The current
architecture associates each accumulator with a group of 4 vector-scalar
registers. This allowed for an implementation that keeps the operating
systems agnostic to the accumulators while exposing them to user code.
In the future, accumulators can be promoted to full architected state
in a fully backwards compatible way.
}

\hl{
Performance-critical kernels of numerical linear algebra
packages, be it traditional (BLIS}~\cite{10.1145/2764454}\hl{,
OpenBLAS}~\cite{xianyi2016openblas}\hl{, MKL}~\cite{MKL}\hl{,
ESSL}~\cite{ESSL}\hl{) or modern (Eigen}~\cite{eigenweb}\hl{,
oneDNN}~\cite{oneDNN}\hl{) ones, are hand-crafted with either compiler
built-ins or in assembly code.  This has enabled the development of
MMA-enabled OpenBLAS and Eigen for immediate consumption while new
compilation techniques are developed.
}

\hl{
The rest of this paper explains in more detail how this
approach, introduced in POWER10, works.
}

\commentout{
\section{Introduction}
\label{Sec:Introduction}

Dense numerical linear algebra is an important class of computations and
the basis of several modern algorithms of great relevance in scientific,
machine learning and business analytics applications. The collection
of building blocks for dense numerical linear algebra has for decades
been standardized in the Basic Linear Algebra Subprograms (BLAS). BLAS
is a \emph{de facto} standard and high performance implementations for
it exist for a broad spectrum of processors in both Open Source (\eg,
BLIS~\cite{10.1145/2764454}, OpenBLAS~\cite{xianyi2016openblas},
Eigen~\cite{eigenweb}) and proprietary (\eg, MKL~\cite{MKL},
ESSL~\cite{ESSL}) libraries.

The highest performing component of the BLAS comprises those routines in the
BLAS3 subset. Whereas BLAS1 and BLAS2 routines include vector-vector and
vector-matrix operations that perform an amount of computation that is
proportional to the size of the data, BLAS3 routines include matrix-matrix
operations that perform $O(n^3)$ computation on $O(n^2)$ data.  This has
enabled various generations of processors to continually improve the
performance of BLAS3 operations while managing the pressure on the
memory subsystem. At any level of the memory hierarchy (from registers
to external storage), the computational intensity of BLAS3 operations
can be improved by increasing the size of storage (and corresponding
problem size).

The key computational kernel of BLAS3 is the innermost loop (also called
the micro-kernel) of matrix multiplication. This micro-kernel computes
a ``squarish'' ($m \times n$), register resident, panel of the output
matrix $C$ as the product of a short and wide ($m \times k, k \gg m$)
input matrix $A$ by a tall and skinny ($k \times n, k \gg n$) input
matrix $B$.  The computation is expressed as a sequence of $k$ rank-$1$
updates to the panel of $C$ by the outer product of an $m$-element column
of $A$ and an $n$-element row of $B$. These rank-$1$ update operations
are computationally intense, since it only takes $m$ elements of $A$
and $n$ elements of $B$ to perform $mn$ multiply-add operations.

The recently released Power ISA\texttrademark\ Version 3.1 has
introduced an entirely new facility: The VSX Matrix-Multiply Assist
(MMA) instructions~\cite{PowerISA3.1}. These instructions directly
implement one or more (depending on the data type) rank-$1$ operations
on small matrices and vectors and can be used to speed up the execution
of critical dense linear algebra kernels.

While developing these rank-$k$ update instructions, it was relatively
straightforward to use the existing vector-scalar register (\vsr s) in
the Power ISA to hold the input columns and rows for matrices $A$ and $B$,
respectively. Holding the result panel of $C$, however, would require
multiple of these vector scalar registers. Therefore, we introduced a
new set of registers called \emph{accumulators}.  Each accumulator is
512 bits wide and can hold a $4 \times 4$ matrix of 32-bit elements. The
new instructions were architected to use an accumulator register for the
(source and target) panel of $C$ and vector-scalar registers for the
inputs $A$ and $B$.

Although there are ongoing activities to develop compilation techniques
that will translate programs written in high-level languages to the
new MMA instructions, we expect most MMA code to be manually generated
in the near future. To facilitate both the development and maintenance
of this hand-written code, we advocate the use of compiler \builtins\
that encapsulate the new operations provided by the MMA facility.

Section~\ref{Sec:ISA} of this paper presents the new MMA facility
architecture, including the new registers and instructions
added to Power ISA.  Section~\ref{Sec:POWER10} discusses the
implementation of the new facility in the announced IBM POWER10
processor~\cite{power10-hotchips2020}.  Section~\ref{Sec:Builtins}
introduces the new compiler \builtins\ that support programming the
new facility. In particular, we discuss two ancillary \builtins\ that
provide the match between the traditional vector-scalar registers of the
Power ISA and the new \emph{accumulator} registers in the MMA facility.
Section~\ref{Sec:Studies} illustrates programming the MMA facility
through various case studies.  Section~\ref{Sec:Performance} presents
performance results that demonstrate the significant gains achieved by the
new MMA facility, including comparisons with the immediate predecessor
POWER9~\cite{power9-ieeemicro2017,power9-core-jrd2018} processor,
whereas Section~\ref{Sec:Power} discusses the power characteristics from
computing with that facility.  Finally, Section~\ref{Sec:Conclusions}
presents our conclusions.
}

\section{Instruction set architecture of MMA}
\label{Sec:ISA}

The MMA facility is fully integrated in the Power ISA. That is, MMA
instructions can appear anywhere in the instruction stream and can
interleave with any other Power ISA instruction. 
They are fetched, decoded and dispatched, like any other instruction,
by the front-end component of the core (called the \emph{Instruction Fetch Unit}, or {\sf IFU}).

MMA instructions are executed by a dedicated functional unit (called
the \emph{Matrix Math Engine}, or \mme), with access to both the Power
ISA vector-scalar registers ($\vsr[0:63]$ -- 64 registers, 128 bits wide
each) and a new set of \emph{accumulator registers}, described below. In
a superscalar, out-of-order processor, such as the future IBM POWER10
processing core, execution of MMA instructions can completely overlap
with the execution of other Power ISA instructions.

\subsection{MMA registers}
\label{Sec:Registers}

\begin{figure}[htb]
    \hrule
    \vspace{-3ex}
    \begin{center}
    \includegraphics[trim=50 150 350 60,clip,width=0.5\textwidth]{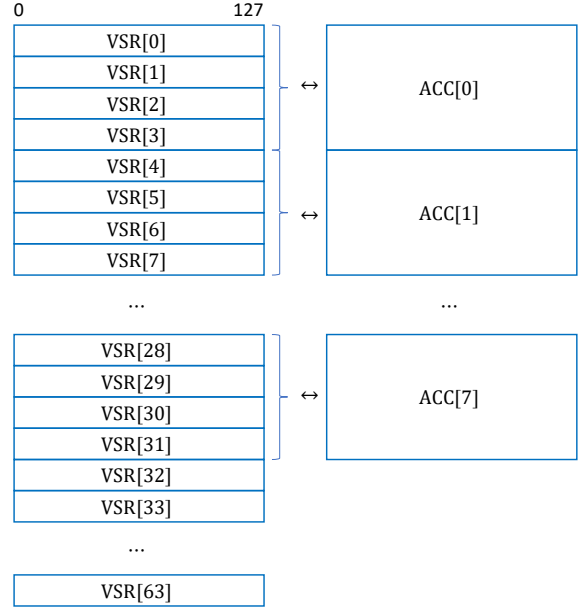}
    \end{center}
    \vspace{-5ex}
    \caption{The MMA facility adds eight 512-bit accumulator registers
    ($\acc[0:7]$) to the Power ISA register set, which includes 64
    vector-scalar registers ($\vsr[0:63]$) of 128 bits each. Each accumulator
    is associated with a group of four consecutive 128-bit vector-scalar
    registers. The $\vsr[32:63]$ vector-scalar registers are not associated
    with, nor do they conflict with, any accumulator.}
    \hrule
\label{Fig:Registers}

\end{figure}

The MMA facility defines a set of eight (8) 512-bit accumulator
registers. Each accumulator register can hold one of three different
kinds of data:
\begin{enumerate}
	\item A $4 \times 2$ matrix of 64-bit double precision
		floating-point elements (\code{fp64}).
	\item A $4 \times 4$ matrix of 32-bit single precision
		floating-point elements (\code{fp32}).
	\item A $4 \times 4$ matrix of 32-bit signed integer
		elements (\code{int32}).
\end{enumerate}
Each of the eight accumulator registers ($\acc[0:7]$) is associated with
a group of four vector-scalar registers (from the $\vsr[0:31]$ subset),
as shown in Figure~\ref{Fig:Registers}.  The architecture requires that,
as long as a particular accumulator register is in use, the associated
four vector-scalar registers must not be used.  Vector-scalar registers
$\vsr[32:63]$ are not associated with any accumulator register and
therefore can always be used while the MMA facility is active.

\subsection{MMA instructions}
\label{Sec:Instructions}

MMA instructions fall into one of three categories:
\begin{enumerate}
	\item \emph{Accumulator Move Instructions}: These instructions
		move data between the accumulator registers and
		their associated vector-scalar registers. 
		(See Table~\ref{Tab:Instructions}(a).)
	\item \emph{Integer rank-$k$ Update Instruction}: These are
		integer arithmetic instructions that update the
		elements of an accumulator with the product of
		input matrices. (See Table~\ref{Tab:Instructions}(b).)
	\item \emph{Floating-point rank-$k$ Update Instructions}:
		These are floating-point arithmetic instructions
		that update the elements of an accumulator with the
		product of input matrices and/or vectors.
		(See Table~\ref{Tab:Instructions}(c).)
\end{enumerate}
The arithmetic instructions (integer and floating-point) have both
a \emph{prefix form}, which always begins with \code{pm}, and
a \emph{conventional form}, without the prefix. We will
start our discussion with the conventional forms of the instructions
and cover the prefix forms later.

\begin{table*}[htb]
    \hrule
	\caption{{\rm MMA instructions. Those instructions with a \code{pm}
	prefix belong to the new class of \emph{prefix instructions}
	in Power ISA, which are 64 bits in size. The others have the
	traditional format (32-bit fixed size). The arithmetic instructions have
	an optional 2-letter suffix that indicates how the product of the input matrices 
	should be added to the target accumulator: \code{pp} -- positive product, positive accumulator,
	\code{np} -- negative product, positive accumulator, \code{pn} -- positive product, negative accumulator,
	\code{nn} -- negative product, negative accumulator. The optional \code{s} suffix indicates
	the  use of saturating arithmetic for the integer instructions.}}
	\begin{center}
	{(a) Accumulator Move Instructions.}
\begin{tabular}{|l|l|}
\hline
	Instruction 		&  Description 																\\
\hline
	\instr{xxsetaccz}	& \begin{minipage}{0.75\textwidth} Set all elements of the target accumulator to 0 \end{minipage} 					\\
	\instr{xxmfacc}		& \begin{minipage}{0.75\textwidth} Move the contents of the source accumulator to the associated vector-scalar registers \end{minipage} \\
	\instr{xxmtacc}		& \begin{minipage}{0.75\textwidth} Move the contents of a group of vector-scalar registers to the associated accumulator \end{minipage} \\
\hline
\end{tabular}

		\vspace{1.0\baselineskip}
	{(b) Integer rank-$k$  update instructions.}
\begin{tabular}{|l|l|}
\hline
	Instruction 		&  Description 																							\\
\hline
	\instr{[pm]xvi16ger2[s][pp]}	& \begin{minipage}{0.75\textwidth} Update a $4 \times 4$ matrix of \code{int32} elements with the product of two $4 \times 2$ matrices of \code{int16} elements  \end{minipage} 	\\
	\instr{[pm]xvi8ger4[pp,spp]}	& \begin{minipage}{0.75\textwidth} Update a $4 \times 4$ matrix of \code{int32} elements with the product of two $4 \times 4$ matrices of \code{int8}/\code{uint8} elements \end{minipage} \\
	\instr{[pm]xvi4ger8[pp]}	& \begin{minipage}{0.75\textwidth} Update a $4 \times 4$ matrix of \code{int32} elements with the product of two $4 \times 8$ matrices of \code{int4} elements \end{minipage} \\
\hline
\end{tabular}

		\vspace{1.0\baselineskip}
	{(c) Floating-point rank-$k$  update instructions.}
\begin{tabular}{|l|l|}
\hline
	Instruction 		&  Description 																\\
\hline
	\instr{[pm]xvbf16ger2[pp,np,pn,nn]}	& \begin{minipage}{0.75\textwidth} Update a $4 \times 4$ matrix of \code{fp32} elements with the product of two $4 \times 2$ matrices of \code{bfloat16} elements  \end{minipage} 	\\
	\instr{[pm]xvf16ger2[pp,np,pn,nn]}	& \begin{minipage}{0.75\textwidth} Update a $4 \times 4$ matrix of \code{fp32} elements with the product of two $4 \times 2$ matrices of \code{fp16} elements  \end{minipage} 	\\
	\instr{[pm]xvf32ger[pp,np,pn,nn]}	& \begin{minipage}{0.75\textwidth} Update a $4 \times 4$ matrix of \code{fp32} elements with the product of two $4$-element vectors of \code{fp32} elements  \end{minipage} 	\\
	\instr{[pm]xvf64ger[pp,np,pn,nn]}	& \begin{minipage}{0.75\textwidth} Update a $4 \times 2$ matrix of \code{fp64} elements with the product of $4/2$-element vectors of \code{fp64} elements  \end{minipage} 	\\
\hline
\end{tabular}

	\end{center}
	\label{Tab:Instructions}
	\hrule
\end{table*}

\subsubsection{Accumulator Move Instructions} The three Accumulator Move
Instructions can be used to initialize the elements of an accumulator
to 0, to move data from vector-scalar registers into an accumulator
register, or to move data from an accumulator register into vector-scalar
registers.  When data is moved from vector-scalar registers into an
accumulator, or when the accumulator elements are initialized to zero,
the accumulator is said to be \emph{primed}. From this point on, the
associated vector-scalar registers should not be used again, until
the accumulator is \emph{deprimed} by moving data from the accumulator
into the associated vector-scalar registers. After a depriming event,
an accumulator should not be used until primed again. (See below for
other instructions that prime accumulators).

\subsubsection{Integer rank-$k$ Update Instructions}
These instructions have the general form
\begin{equation}
	A \leftarrow X Y^T [+ A]
\end{equation}
where $A$ is an accumulator register, holding a $4 \times 4$ matrix
of \code{int32} elements and $X$ and $Y$ are vector-scalar registers,
that must not overlap the accumulator, holding matrices of either 16-,
8- or 4-bit integers. (The bracketed term $[+A]$ is optional and $Y^T$
denotes the transpose of $Y$.) The exact shape of the input matrices
depends on the type of input data, which also defines the value of $k$
in the rank-$k$ update operation.

Vector-scalar registers in Power ISA are always 128 bits wide.
Therefore, when the input data are 16-bit integers (\code{int16}),
the $X$ and $Y$ registers are interpreted as $4 \times 2$ matrices, so
that $X Y^T$ produces a $4 \times 4$ matrix as a result. That product
matrix can be optionally added to the current value of the accumulator
or directly stored in the accumulator. Instructions that simply write
the value of $X Y^T$ into the target accumulator automatically prime that
accumulator. The accumulation form of the instructions, with the \code{pp}
suffix in case of integer types, require that the target accumulator be
previously primed with an initial value.

When using 16-bit integer inputs, with the \instr{xvi16ger2} instructions,
there are two choices for the arithmetic model: the more conventional
\emph{modulo} arithmetic, where the largest representable integer is
followed by the smallest representable integer, and \emph{saturating}
arithmetic model, where adding positive values to the largest
representable integer or negative values to the smallest representable
integer does not change the target value. Instructions with the \instr{s}
suffix (\eg, \instr{xvi16ger2s}) use the saturating model.

For 8-bit integer inputs, with the \instr{xvi8ger4} instructions, the $X$
and $Y$ registers are interpreted as $4 \times 4$ matrices. Whereas the
contents of $X$ are a $4 \times 4$ matrix of signed 8-bit integer elements
(\code{int8}), the contents of $Y$ are a $4 \times 4$ matrix of unsigned
8-bit integer elements (\code{uint8}).  This mixing of signed and unsigned
8-bit integer inputs has been common practice since early generations
of vector instructions, and is also present in modern deep learning
libraries~\cite{MKL-DNN}.  As with the 16-bit inputs case, the product
can be optionally added to the current value of the accumulator. The
same requirements regarding automatic priming of the target accumulator
also hold.

The \instr{xvi8ger4} instructions offer the same choice of modulo
\emph{vs} saturating arithmetic as the \instr{xvi16ger2} instructions.
Saturating arithmetic is only available in the accumulation-form of
the instruction (suffix \code{spp}), since a product of $4 \times 4$
8-bit matrices cannot overflow a 32-bit integer result.

The final family of integer rank-$k$ update instructions consist of the
\instr{xvi4ger8} instructions. In this case, the $X$ and $Y$ registers
are interpreted as $4 \times 8$ matrices of signed 4-bit integer elements
(\code{int4}). The product $XY^T$ can be optionally added to the contents
of the target accumulator (suffix \code{pp}). It is unlikely for a sum
of products of 4-bit inputs to overflow a 32-bit accumulator. Therefore,
only a modulo arithmetic version of this operation is provided.

\subsubsection{Floating-point rank-$k$ Update Instructions}
These instructions have the general form
\begin{equation}
	A \leftarrow [-] X Y^T [{\pm} A]
\end{equation}
where $A$ is an accumulator register, holding either a $4 \times 2$
matrix of double-precision (\code{fp64}) elements or a $4 \times 4$
matrix of single-precision (\code{fp32}) elements.  $X$ and $Y$ are
vector-scalar registers, that must not overlap the accumulator, holding
matrices or vectors of 16-, 32- or 64-bit floating-point values. (In
one case discussed below, $X$ is a pair of vector-scalar registers.)
The product of the input matrices or vectors can be optionally negated,
and then added or subtracted to the current contents of the target
accumulator.  The optional \code{pp}, \code{np}, \code{pn}, and \code{nn}
suffixes control the accumulation operation.
The first \code{p/n} in the suffix specifies either a positive or negated
product and the second \code{p/n} specifies either a positive or negated
accumulator.

There are two families of rank-2 update instructions for 16-bit input
elements. The \instr{xvbf16ger2} instructions treat the inputs in
\emph{brain float 16} format (\code{bf16})~\cite{8342167}, whereas the
\instr{xvf16ger2} instructions treat the inputs in IEEE half-precision
format (\code{fp16})~\cite{8766229}.  In both cases, $X$ and $Y$ are $4
\times 2$ matrices of 16-bit elements, producing a $4 \times 4$ matrix
product (optionally negated) that can then be added to the (optionally
negated) target accumulator.  Just as with the integer rank-$k$ update
instructions, the nonaccumulation-form of the floating-point instructions
automatically prime the target accumulator.

For 32-bit inputs, the \instr{xvf32ger} instructions use $X$ and $Y$ as
4-element vectors of single-precision (\code{fp32}) values, computing
a $4 \times 4$ outer-product that can then be optionally negated and
added to the (optionally negated) target accumulator.

The double-precision instructions \instr{xvf64ger} break the usual
conventions for the rank-$k$ update instructions. First, the accumulator
is treated as a $4 \times 2$ matrix of double-precision elements
(\code{fp64}). The $X$ input is a $4$-element vector of \code{fp64} values
(consisting of an even-odd pair of adjacent vector-scalar registers)
and the $Y$ input is a $2$-element vector of \code{fp64} values.
None of the input vector-scalar registers can overlap the accumulator.
The $X Y^T$ outer-product is computed, producing a $4 \times 2$ result.
That result is optionally negated and added to the (optionally
negated) accumulator.

\subsection{Prefixed instructions}

Power ISA\texttrademark\ Version 3.1 introduces \emph{prefixed
instructions}.  Whereas all Power ISA instructions pre-dating Version
3.1 consist of a single 32-bit word encoding, prefixed instructions are
64 bits long, consisting of a 32-bit prefix word followed by a 32-bit
suffix word.

Each of the integer and floating-point rank-$k$ update instructions in the
MMA facility has a prefix version that extends the previously discussed
functionality of the base instruction.  That extended functionality
consists of immediate \emph{mask} fields that specify the exact rows
of $X$ and columns of $Y^T$ to be used in the computation. When the MMA
instruction is of rank 2 or higher ($k \geq 2$), a third \emph{product
mask} field can specify the exact outer products to use when computing
the result.

The masking feature of the prefixed variants is better illustrated with
an example. Consider the multiplication of two $4 \times 2$ matrices $X$
and $Y$ of half-precision floating-point elements (\code{fp16}) through
the instruction
\begin{quotation}
	\code{pmxvf16ger2pp $A$, $X$, $Y$, $x$, $y$, $p$}
\end{quotation}
where $A$ is the accumulator, $x$ and $y$ are the 4-bit immediate fields
specifying the masks for input matrices $X$ and $Y$ respectively, and $p$
is the 2-bit immediate field specifying the product mask. Let $x = x_0
x_1 x_2 x_3$, $y = y_0 y_1 y_2 y_3$ and $p = p_0 p_1$, where the $x_i$,
$y_j$ and $p_k$ are single bit values (0 or 1).  The resulting value of
each element $A_{ij}$ of accumulator $A$ is computed by
\begin{equation}
	A_{ij} \leftarrow \sum_{k=0,1} [(p_k(x_i X_{ik} \times y_j Y_{jk})] + A_{ij}. 
\end{equation}
In other words, the $x$ mask enables/disables rows of $X$, the $y$ mask
enables/disables columns of $Y^T$ and the $p$ mask enables/disables
the specific partial products along the inner dimension (the $k$ from
rank-$k$) of the matrix multiply.  Computations on disabled rows and
columns are not performed and, therefore, exceptions are not generated
for those computations.

The prefix variant of the rank-$k$ update instructions can be used
to compute operations on matrices of shape different than the shape
directly supported by the conventional instructions. This can be
useful when computing residual loop iterations after a matrix is
blocked into multiples of the default size.  For the \code{xvf32ger} and
\code{xvf64ger} families of instructions, only the $x$ and $y$ masks can
be specified, since the rank of those operations is always one ($k = 1$).

\section{Implementation in the POWER10 processor}
\label{Sec:POWER10}

\begin{figure}
    \hrule
    \vspace{-0.2ex}
    \begin{center}
    \includegraphics[trim=60 110 50 80,clip,width=0.50\textwidth]{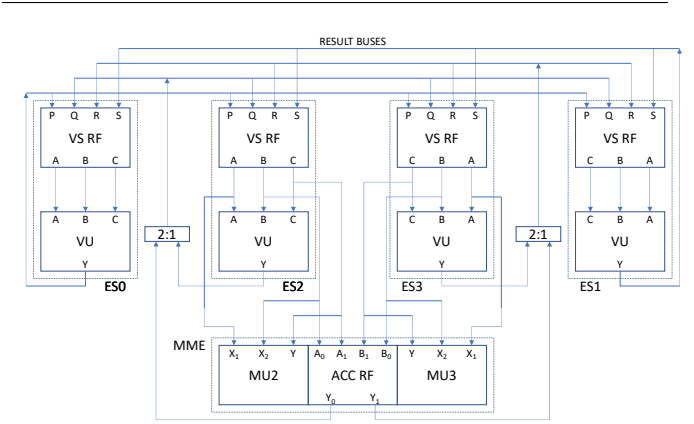}
    \end{center}
    \vspace{-2.5ex}
    \caption{Block diagram of how the matrix math unit interacts
	with the rest of the POWER10 core. All buses shown are 128 bits wide.}
    \label{Fig:Micro}
    \hrule
\end{figure}

Figure~\ref{Fig:Micro} is a block diagram for the backend of the POWER10
core. Only a subset of the backend, relevant to the execution of MMA
instructions, is illustrated.  We do not cover aspects of the execution
of either scalar instructions or load/store instructions, which are not
relevant to the execution of matrix instructions.

The backend consists of four (4) execution slices (\micro{ES}$[0:3]$) and
an attached matrix math engine (\micro{MME}). An execution slice contains
a register file (\micro{VS RF}) for the 128-bit wide vector-scalar
registers and an execution unit (\micro{VU}) for performing operations on
those registers.  On a given cycle, each slice can issue one instruction
for execution. Slices 2 and 3 can issue either a vector instruction or
an MMA instruction.  Slices 0 and 1 can only issue vector instructions.

The matrix math engine logically consists of two (2) execution pipelines
(\micro{MU2} and \micro{MU3} for instructions issued from slices 2 and
3, respectively) sharing a common accumulator register file (\micro{ACC
RF}). This organization supports the execution of two rank-$k$ update
instructions per cycle.

\begin{figure}
    \hrule
    \vspace{-0.5ex}
    \begin{center}
    \includegraphics[trim=150 80 170 70,clip,width=0.50\textwidth]{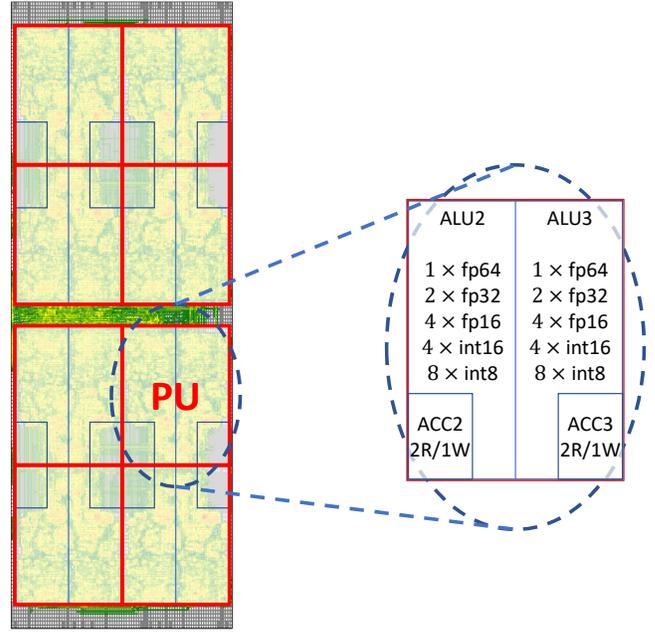}
    \end{center}
    \vspace{-3.0ex}
    \caption{Each matrix unit consists of a $4 \times 2$ grid of basic tiles.
	Each tile has a 64-bit slice of the accumulator register file (each accumulator
	is 512 bits wide) and the corresponding multiply-add functional units
	to operate on that data slice.}
    \label{Fig:Tiles}
    \hrule
\end{figure}

The physical implementation of the \mme\ is shown in
Figure~\ref{Fig:Tiles} and consists of a $4 \times 2$ grid of processing
units (\micro{PU}).  Each processing unit has two identical halves, one
for each of the issuing slices (2 and 3).  Each half includes a 64-bit
slice of the accumulator register file (\micro{ACC2} and \micro{ACC3})
with two read and one write ports.  The arithmetic and logic unit in each
half (\micro{ALU2} and \micro{ALU3}) can perform one double-precision or
two single-precision floating-point multiply-add(s).  It can also perform
4, 8, or 16 multiply-adds of 16-, 8-, or 4-bit data, respectively. The
result of each ALU is always written to the corresponding accumulator
register file but the input can come from either one. (Hence the
requirement for two read ports.)

Whereas the accumulators (both input and output) of each instruction
are stored in the accumulator register file, the $X$ and $Y$ inputs are
sourced from two of the vector-scalar register files.  The fetch buses
from the vector-scalar register files bring $X$ and $Y$ operands for
the outer product instructions and transfer data from the vector-scalar
registers to the accumulators.  The result buses transfer data from
accumulators to the vector-scalar registers.  It takes two (2) cycles
to transfer four (4) vector-scalar registers to an accumulator and four
(4) cycles to transfer one accumulator to 4 vector-scalar registers. Up
to two transfers can be performed simultaneously.

During the computation phase of a math kernel, the accumulator data
stays local to the matrix math engine. Only the $X$ and $Y$ inputs have
to be brought from the register files. Furthermore, no output is placed
on the results buses. This leads to a more power efficient execution of
those kernels.
 
We compare the current MMA approach using the POWER10 matrix math
engine (MME) with two other alternatives to improving the performance
of processors for dense numerical linear algebra kernels: (1) expanding
the vector width and (2) building a dedicated matrix-multiply unit.

The more conventional approach of widening the vector registers and
vector units has been adopted by several products~\cite{8641463,9229611,A64fx}. When comparing it to
the approach adopted in the matrix math facility:
\begin{enumerate}
    \item The new matrix math facility instructions have no impact to the
    rest of the architecture (vector registers and vector instructions
    stay exactly the same) and minimal impact to the micro-architecture
    (the matrix math engine is attached to the execution slices, which do
    not require any change in supporting their operations.)  In contrast,
    widening vectors would require deeper architectural changes, either
    in the form of new instructions~\cite{AVX-512} or switching to a
    scalable vector approach~\cite{Stephens_2017}, wider vector registers,
    and wider vector units.

    \item When performing a $4 \times 4$ outer-product of single-precision
    (32-bit) floating-point data, only $2 \times 128$-bit vector registers
    have to be transmitted from the register file to the matrix math
    engine. The much larger $512$-bit accumulator resides entirely within
    the matrix math engine. A comparable $512$-bit wide vector unit
    would require $3 \times 512$-bit registers to be fetched and one to be
    written back to the register file for the same 16 (single-precision)
    floating-point multiply-add operations.

    \item The physical design of the matrix math engine has a natural
    two-dimensional layout (see Figure~\ref{Fig:Tiles}) that follows
    the structure of the outer-product computation (either $4 \times 2$
    for double-precision or $4 \times 4$ for narrower data types). Vector
    computations are naturally one-dimensional and may need additional
    constructs to fold into two-dimensional arrangements.

    \item The outer product is a BLAS2 operation and the natural
    algorithmic operation for the most important dense numerical linear
    algebra kernels. It is directly supported by the instructions of
    the matrix math facility. In comparison, processors with vector
    instructions require additional steps to transform a two-dimensional
    BLAS2 outer product into one-dimensional BLAS1 operations that are
    supported by the vector instructions. Those additional operations
    can include broadcast loads or splat instructions.

    \item The issue-to-issue latency for the matrix math facility
    instructions is reduced when compared to comparable vector
    instructions, since the accumulators are already in the functional
    unit, as opposed to vector registers that are fetched from a separate
    register file.
\end{enumerate}

Another emerging approach is to build a dedicated matrix multiply unit, either
to the core or at the chip level. This solution compares to the
matrix math facility as follows:
\begin{enumerate}
    \item The instructions of the matrix math facility are part of the
    instruction stream of a thread and much finer grain than a complete
    matrix multiplication.

    \item The instructions of the matrix math facility can be used
    as building blocks of other computations, such as convolution,
    triangular solve and discrete Fourier transform.
\end{enumerate}

\section{Programming the MMA facility}
\label{Sec:Builtins}

Generation of MMA facility code from high-level language constructs is
an active area of research. Currently, most code that uses those new
instructions is manually generated, with explicit invocation of the
new operations. While directly programming in assembly instructions is
always an option for MMA exploitation, we advocate the use of compiler
\builtins\ as a preferred alternative~\cite{MMABestPractices}.

\Builtins\ are functions with pre-defined semantics, known to the
compiler. The compiler can directly emit code for these \builtin\
functions, and quite often they translate one-to-one to native
machine instructions. They represent a compromise in abstraction. The
programmer has detailed control of the operations performed by the
machine while implementation of the \builtins\ in the compiler can
choose to include additional semantics about the instructions. This
additional information can then be used throughout the compilation process
to enable optimizations. Furthermore, low-level optimizations such as
instruction scheduling and register allocation are left to the compiler.

The open source GNU Compiler Collection (GCC), starting with version 10.2,
has already been augmented with \builtins\ for the MMA facility, while
work on Clang/LLVM compilers is under way. This is in addition to the
various \builtins\ that were already implemented, including architecture
agnostic and Power ISA-specific \builtins. This provides performance
and functional portability across the different compilers. For this
reason, and the simplicity compared with direct assembly programming,
we believe programming with \builtins\ is the preferred approach for
broader exploitation of the MMA facility.

MMA \builtins\ make use of three data types to specify data
manipulated by those \builtins:
\begin{quote}
    \code{\_\_vector unsigned char} -- a 16-byte vector, used for most rank-$1$ update operations \\
    \code{\_\_vector\_pair}	    -- a 32-byte vector, used for the \code{fp64} rank-$k$ update operations \\
    \code{\_\_vector\_quad}	    -- a 64-byte accumulator
\end{quote}

The new MMA \builtins\ are summarized in Table~\ref{Tab:Builtins}.
Most \builtins\ correspond one-to-one to machine instructions, as
shown in the table. Two of the \builtins\ provide an ancillary role
to the compiler, by constructing accumulators from vectors and
extracting vectors from accumulators.

\begin{table*}[htb]
    	\hrule	
	\caption{{\rm MMA \builtins. Each MMA instruction has a corresponding \builtin\ function with
	pre-defined semantics known to the compiler. By programming with \builtins, the programmer
	can specify the exact operations to be performed by the hardware, while
	leaving register allocation and instruction scheduling to the compiler.
	In the table below, $A$ represents an accumulator (and $\&A$ its address), 
	where $x$, $y$, $z$ and $t$ are vectors. $Q$ are vector pairs, used to hold a
	4-element vector of \code{fp64} values. Finally, \code{u2}, \code{u4} and \code{u8}
	are 2-, 4- and 8-bit unsigned integer literals used to define the masks in
	the prefixed instructions.}}
	\label{Tab:Builtins}
	\begin{center}
	    \vspace{-2ex}
{\scriptsize
\begin{tabular}{|l|l|}
	\hline
	Instruction 				&  \builtin 									\\
	\hline
						& \code{\_\_builtin\_mma\_assemble\_acc($\&A$,$x$,$y$,$z$,$t$)}			\\	
						& \code{\_\_builtin\_mma\_disassemble\_acc($\&x$,$\&A$)}			\\
	\instr{xxsetaccz}			& \code{\_\_builtin\_mma\_xxsetaccz($\&A$)} 					\\
	\instr{xxmfacc}				& \code{\_\_builtin\_mma\_xxmfacc($\&A$)}   		 			\\
	\instr{xxmtacc}				& \code{\_\_builtin\_mma\_xxmtacc($\&A$)}       				\\
	\instr{xvi16ger2[s][pp]}		& \code{\_\_builtin\_mma\_xvi16ger2[s][pp]($\&A$,$x$,$y$)} 			\\
	\instr{pmxvi16ger2[s][pp]}		& \code{\_\_builtin\_mma\_pmxvi16ger2[s][pp]($\&A$,$x$,$y$,u4,u4,u2)} 		\\
	\instr{xvi8ger4[pp,spp]}		& \code{\_\_builtin\_mma\_xvi8ger4[pp,spp]($\&A$,$x$,$y$)}  			\\
	\instr{pmxvi8ger4[pp,spp]}		& \code{\_\_builtin\_mma\_pmxvi8ger4[pp,spp]($\&A$,$x$,$y$,u4,u4,u4)} 		\\
	\instr{xvi4ger8[pp]}			& \code{\_\_builtin\_mma\_xvi4ger8[ppp]($\&A$,$x$,$y$)}  			\\
	\instr{pmxvi4ger8[pp]}			& \code{\_\_builtin\_mma\_pmxvi4ger8[pp]($\&A$,$x$,$y$,u4,u4,u8)} 		\\
	\instr{xvbf16ger2[pp,np,pn,nn]}		& \code{\_\_builtin\_mma\_xvbf16ger2[pp,np,pn,nn]($\&A$,$x$,$y$)}  		\\
	\instr{pmxvbf16ger2[pp,np,pn,nn]}	& \code{\_\_builtin\_mma\_pmxvbf16ger2[pp,np,pn,nn]($\&A$,$x$,$y$,u4,u4,u2)}	\\
	\instr{xvf16ger2[pp,np,pn,nn]}		& \code{\_\_builtin\_mma\_xvf16ger2[pp,np,pn,nn]($\&A$,$x$,$y$)}  		\\
	\instr{pmxvf16ger2[pp,np,pn,nn]}	& \code{\_\_builtin\_mma\_pmxvf16ger2[pp,np,pn,nn]($\&A$,$x$,$y$,u4,u4,u2)} 	\\
	\instr{xvf32ger[pp,np,pn,nn]}		& \code{\_\_builtin\_mma\_xvf32ger[pp,np,pn,nn]($\&A$,$x$,$y$)}  		\\
	\instr{pmxvf32ger[pp,np,pn,nn]}		& \code{\_\_builtin\_mma\_pmxvf32ger[pp,np,pn,nn]($\&A$,$x$,$y$,u4,u4)} 	\\
	\instr{xvf64ger[pp,np,pn,nn]}		& \code{\_\_builtin\_mma\_xvf64ger[pp,np,pn,nn]($\&A$,$Q$,$y$)}  		\\
	\instr{pmxvf64ger[pp,np,pn,nn]}		& \code{\_\_builtin\_mma\_pmxvf64ger[pp,np,pn,nn]($\&A$,$Q$,$y$,u4,u2)} 	\\
	\hline
\end{tabular}
}

	\end{center}
	\hrule
\end{table*}

The \code{\_\_builtin\_mma\_assemble\_acc} performs a \emph{gather}
operation, collecting four 16-byte vectors $x$, $y$, $z$, and $t$
into an accumulator $A$.  At first glance, this \builtin\ may seem
identical to the \instr{xxmtacc} instruction but that instruction
(and the corresponding \code{\_\_builtin\_mma\_xxmtacc} \builtin) only
transfers data between an accumulator and its corresponding vector-scalar
registers, whereas the \code{\_\_builtin\_mma\_assemble\_acc} \builtin\
can initialize an accumulator from any set of four vectors.

Similarly, the \code{\_\_builtin\_mma\_disassemble\_acc} \builtin\
performs a \emph{scatter} operation, extracting the contents of
an accumulator into an array of vectors that can then be used
individually in the code.  This is different than the transfer
accomplished by the \instr{xxmfacc} instruction (and corresponding
\code{\_\_builtin\_mma\_xxmfacc} \builtin).  We give an illustration
of using the \code{\_\_builtin\_mma\_disassemble\_acc} \builtin\ in
Figure~\ref{Fig:MacrosDGEMM}.

When programming with \builtins, there are some general guidelines to
follow in order to help the compiler generate good quality code.  First,
it is not advisable to explicitly use the \code{\_\_builtin\_mma\_xxmfacc}
and \code{\_\_builtin\_mma\_xxmtacc} \builtins.  Although they are
provided for completeness, it is better to simply provide the compiler
with the list of vectors to initialize an accumulator with, using the
\code{\_\_builtin\_mma\_assemble\_acc} \builtin.  Correspondingly, it is
better to just have the compiler decompose an accumulator into a group
of vectors, using the \code{\_\_builtin\_mma\_disassemble\_acc} \builtin.

Second, there are limitations to passing accumulators across function
calls, and even when supported it is likely to cause a performance
degradation.  A possible exception to this guideline is when one can
be certain the compiler will inline the function, and therefore remove
superfluous copies. (This is a common practice in C++ template libraries.)
For most cases, the programmer should limit accumulator usage to within
a function and avoid having function calls while using accumulators.

Third, the programmer must be conscious of the actual number of
accumulators supported by the architecture (8) and not create too many
live accumulator objects in a function. Otherwise, the compiler may be
forced to spill extra accumulators to and from memory, which also causes
a performance degradation.

Finally, and this is more a rule than a guideline, the programmer must
not use an accumulator that has not been primed. Accumulators can be
primed either by the \code{\_\_builtin\_mma\_assemble\_acc} \builtin,
by the \code{\_\_builtin\_mma\_xxsetaccz} \builtin, or by any of the
nonaccumulating arithmetic rank-$k$ operations.

\section{Case studies}
\label{Sec:Studies}

We present two case studies to illustrate the use of the matrix math
instructions on computations. The first case study is
for the most natural application: matrix multiplication, in this case
of double-precision floating point values ({\sf DGEMM}). The second
case study is in the computation of two-dimensional convolutions,
often used in deep learning, with single-precision floating-point data
({\sf SCONV}).

\subsection{{\sf DGEMM}}
\label{Sec:DGEMM}

\dgemm\ is the general matrix-multiply routine from BLAS, computing
\begin{equation}
    C \leftarrow \alpha A^{[T]} B^{[T]} + \beta C
\end{equation}
where $A$, $B$, $C$ are matrices and $\alpha$, $\beta$ are scalars, all of
type double-precision floating-point.  We consider here only the innermost
kernel found in high-performance libraries~\cite{10.1145/1356052.1356053}.

The inner-most kernel of \dgemm\ computes a register-contained
$m \times n$ block of matrix $C$ as the product of a $m \times k$
block of matrix $A$ and a $k \times n$ block of matrix $B$.
Typically, $k \gg m$ and $k \gg n$, to help amortize the cost
of loading and storing the $C$ block into/from registers.

\begin{figure*}
    \hrule
    \vspace{-3.5ex}
    \begin{tabular}{cc}
	\includegraphics[trim=100 20 100 30,clip,width=0.45\textwidth]{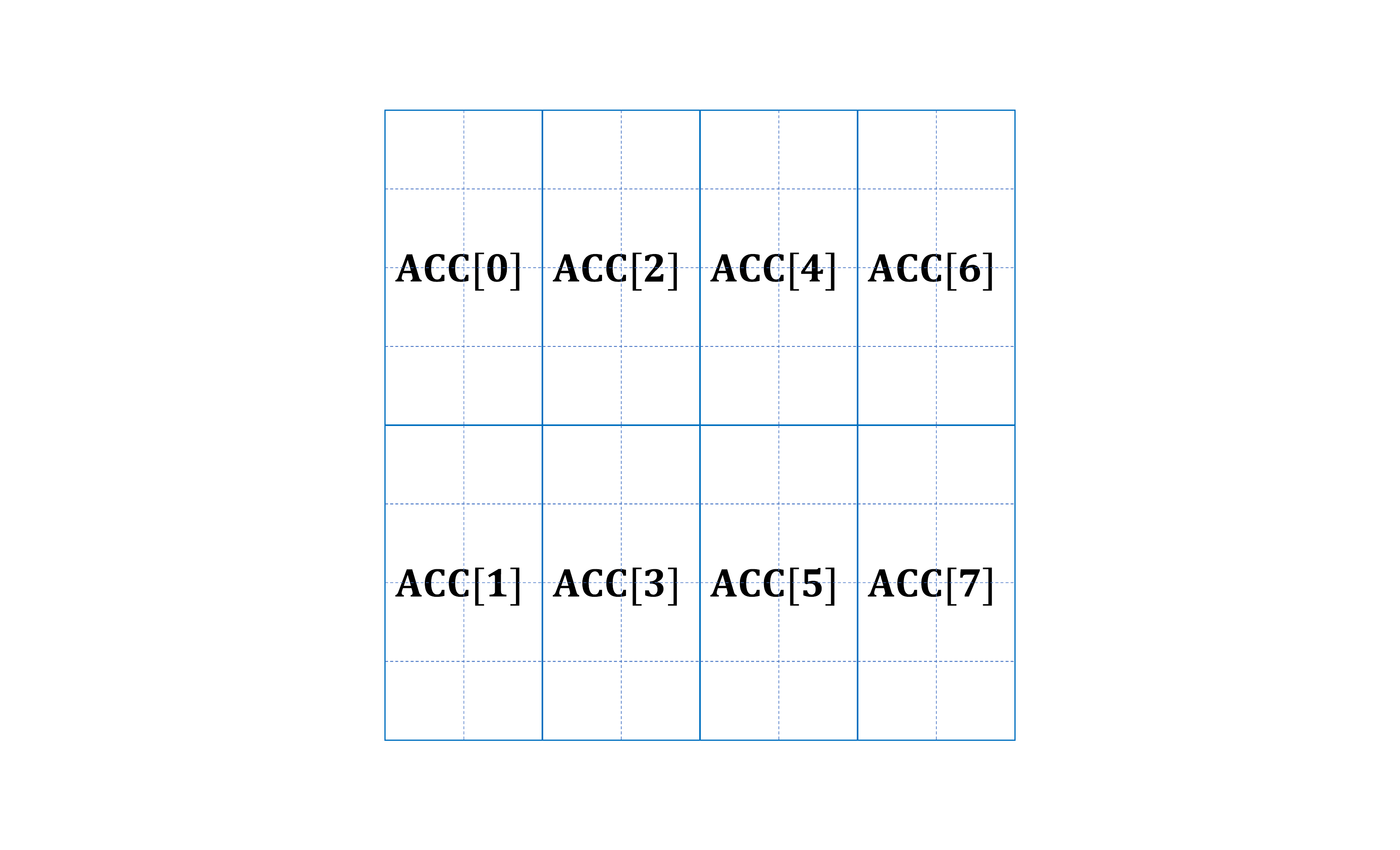} &
	\includegraphics[width=0.45\textwidth]{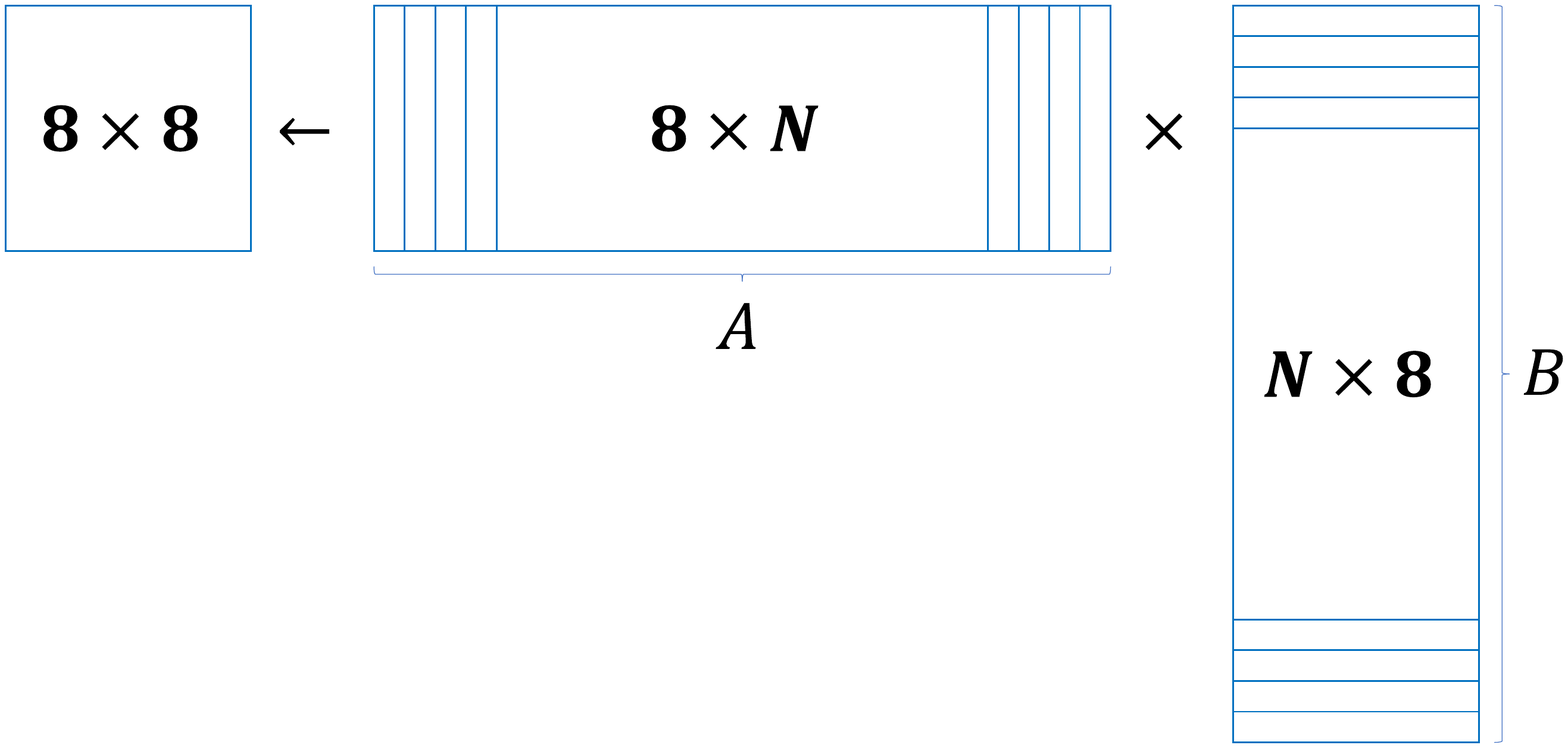} \\
	(a)	& (b)
    \end{tabular}
    \vspace{-1.5ex}
    \caption{The \dgemm\ kernel uses all architected accumulators
	to create a virtual $8 \times 8$ accumulator of double-precision elements (a).
	The accumulator is used to compute the product of an $8 \times N$ matrix $A$ and an
	$N \times 8$ matrix $B$ (b).}
    \label{Fig:Accumulator8x8}
    \hrule
\end{figure*}

For our example, we will use all eight architected accumulators
to create a virtual $8 \times 8$ accumulator of double-precision
elements, as shown in Figure~\ref{Fig:Accumulator8x8}~(a).
The accumulator numbers in the figure are for illustration
purpose only. Since we are programming with \builtins, we cannot
control the precise allocation of registers. And that is
not important either. The compiler is free to choose the 
particular allocation that guarantees correctness and will not affect performance.

\subsubsection{The code with \builtins}

Supporting definitions to make the example code more compact are shown
in Figure~\ref{Fig:MacrosDGEMM}.  Lines 1--3 redefine the data types
directly supported by the compilers to names that are more related to
the computation: A 16-byte vector data type (\code{\_\_vector unsigned
char}) is used to represent a two-element vector of double-precision
floating-point numbers (\code{fp64\_2}), whereas a pair of vectors
(\code{\_\_vector\_pair}) represents a four-element vector of
double-precision floating-point numbers (\code{fp64\_4}) and a group
of four vectors (\code{\_\_vector\_quad}) represents a $4 \times 2$
matrix of double precision floating-point numbers (\code{fp64\_4x2}).

\begin{figure}
    \hrule
    {\scriptsize
\begin{lstlisting}[language=C,firstline=29,lastline=58,escapechar=`,numbers=left,numbersep=2pt]
typedef	__vector_quad		fp64_4x2;
typedef	__vector_pair		fp64_4;
typedef __vector unsigned char	fp64_2;

#define mma_store_acc(AS, A, D)			\
{						\
    fp64_2	a[4];				\
    __builtin_mma_disassemble_acc(a,&(AS));	\
    *((fp64_2*)A+D+0) = a[0];			\
    *((fp64_2*)A+D+1) = a[1];			\
    *((fp64_2*)A+D+2) = a[2];			\
    *((fp64_2*)A+D+3) = a[3];			\
}

#define mma_xvf64_8x8(acc, op, X, Y)			\
{							\
    fp64_4	x0, x1;					\
    fp64_2	y0, y1, y2, y3;				\
    x0 = *((fp64_4*)X+0); x1 = *((fp64_4*)X+1);		\
    y0 = *((fp64_2*)Y+0); y1 = *((fp64_2*)Y+1); 	\
    y2 = *((fp64_2*)Y+2); y3 = *((fp64_2*)Y+3); 	\
    __builtin_mma_xvf64 ## op (&(acc[0]), x0, y0);	\
    __builtin_mma_xvf64 ## op (&(acc[1]), x0, y1);	\
    __builtin_mma_xvf64 ## op (&(acc[4]), x1, y0);	\
    __builtin_mma_xvf64 ## op (&(acc[5]), x1, y1);	\
    __builtin_mma_xvf64 ## op (&(acc[2]), x0, y2);	\
    __builtin_mma_xvf64 ## op (&(acc[3]), x0, y3);	\
    __builtin_mma_xvf64 ## op (&(acc[6]), x1, y2);	\
    __builtin_mma_xvf64 ## op (&(acc[7]), x1, y3);	\
}
\end{lstlisting}
    }
    \vspace{-2ex}
    \caption{Supporting definitions for the {\sf DGEMM} kernel code.}
    \label{Fig:MacrosDGEMM}
    \hrule
\end{figure}

Lines 5--13 define macro \code{mma\_store\_acc}, which stores accumulator
\code{AS} in a 64-byte memory location beginning at $\code{D} \times 16$ bytes
past the address in pointer \code{A}.  The accumulator is
first transferred to an array of four 2-element vectors (through the
\builtin\ \code{\_\_builtin\_mma\_disassemble\_acc}) and then the elements
are stored at consecutive 16-byte chunks of memory.

Lines 15--30 of Figure~\ref{Fig:MacrosDGEMM} define macro \code{mma\_xvf64\_8x8}
which computes the outer-product of two 8-element vectors of double
precision floating-point numbers (\code{X} and \code{Y}), accumulating
the result into an $8 \times 8$ accumulator (\code{acc}) represented as an array of
eight $4 \times 2$ accumulators.  The exact operation (accumulating or
not, inverting signs or not) is specified by the \code{op} argument,
which must be one of \code{ger}, \code{gerpp}, \code{gernp}, \code{gerpn},
or \code{gernn}.

The kernel function that computes the product $X Y^T$ of
two $8 \times N$ double-precision matrices $X$ and $Y$ is shown
in Figure~\ref{Fig:KernelDGEMM}. Line 9 tests for an empty multiply. Line
11 declares the array of $4 \times 2$ accumulators that implement the
virtual $8 \times 8$ accumulator.  Line 13 is the initial multiply without
accumulation, which initializes the $8 \times 8$ accumulator. Lines 15-19
are the main loop, which performs the remaining $N-1$ outer-products,
with accumulation.  Finally, lines 21-28 store the components of the $8
\times 8$ accumulator into the result matrix \code{A}. (The layout is
not conventional. That is handled in other layers of \dgemm.)

\begin{figure}
    \hrule
    {\scriptsize
\begin{lstlisting}[language=C,firstline=60,lastline=88,escapechar=`,numbers=left,numbersep=2pt]
void dgemm_kernel_8xNx8
(
    double		*A,
    const double	*X,
    const double 	*Y,
    const uint64_t	 n
)
{
    if (n == 0) return;

    fp64_4x2	acc[8];

    mma_xvf64_8x8(acc, ger, X, Y);

    for (uint64_t i=1; i<n; i++)
    {
	X += 8; Y += 8;
	mma_xvf64_8x8(acc, gerpp, X, Y);
    }

    mma_store_acc(acc[0], A, 0);
    mma_store_acc(acc[1], A, 4);
    mma_store_acc(acc[2], A, 8);
    mma_store_acc(acc[3], A,12);
    mma_store_acc(acc[4], A,16);
    mma_store_acc(acc[5], A,20);
    mma_store_acc(acc[6], A,24);
    mma_store_acc(acc[7], A,28);
}
\end{lstlisting}
    }
    \vspace{-2ex}
    \caption{{\sf DGEMM} $8 \times N \times 8$ kernel code.}
    \label{Fig:KernelDGEMM}
    \hrule
\end{figure}

\subsubsection{The generated machine code}

We compile the source code of Figure~\ref{Fig:KernelDGEMM} using
\code{g++} version $11.0$ in the IBM Advance Toolchain version 15.0
(alpha)~\cite{IBMAT}, with the compiler flags
\begin{quote}
    \code{-mcpu=power10 -O3}
\end{quote}
which explicitly enable MMA support and higher levels of optimization.

The management of the new accumulator registers
present significant challenges to their enablement in compilers.
In particular, the need to transfer data to and from accumulators,
and their overlap with existing vector-scalar registers,
force the compiler to insert various register spill and copy
operations in the intermediate representation of the code.
Those are successfully removed with higher levels of optimization,
which therefore are crucial to get good quality
code from \builtins.

\commentout{
Figure~\ref{Fig:Init} shows the object code generated for source lines
9--13 in Figure~\ref{Fig:KernelDGEMM}. The various \code{sftd} machine
instructions (lines 10--21 and 24--29 of Figure~\ref{Fig:Init} save
the contents of the non-volatile registers that will be modified by the
\dgemm\ kernel. These values are restored at function exit, as discussed
below.  The 4-element vector variables \code{x0} and \code{x1} in line 17
of Figure~\ref{Fig:MacrosDGEMM} are assigned register pairs $\vsr[0:1]$
and $\vsr[32:33]$, respectively, and loaded through 32-byte load
instructions in lines 3-4 of Figure~\ref{Fig:Init}.  Correspondingly, the
2-element vector variables \code{y0}, \code{y1}, \code{y2} and \code{y3}
in line 18 of Figure~\ref{Fig:MacrosDGEMM} are loaded through 16-byte load
instructions in lines 6--9 of Figure~\ref{Fig:Init}.  The $8 \times 8$
initializing outer-product in line 13 of Figure~\ref{Fig:KernelDGEMM}
is implemented by 8 \code{xvf64ger} instructions, in lines 22,23,30--35
of Figure~\ref{Fig:Init}. (Even though the compiler is using some
low-numbered vector-scalar registers as inputs, those registers are
defined and used before any conflicting accumulator is primed.)
 
\begin{figure}[h]
    \hrule
    {\scriptsize
\begin{lstlisting}[language=C,firstline=1142,lastline=1181,escapechar=`,numbers=left,numbersep=0pt]
    100016b0:	00 00 26 2c 	cmpdi   r6,0
    100016b4:	20 00 82 4d 	beqlr
    100016b8:	00 00 04 18 	lxvp    vs0,0(r4)
    100016bc:	20 00 24 18 	lxvp    vs32,32(r4)
    100016c0:	01 00 26 28 	cmpldi  r6,1
    100016c4:	01 00 45 f5 	lxv     vs10,0(r5)
    100016c8:	11 00 65 f5 	lxv     vs11,16(r5)
    100016cc:	21 00 65 f4 	lxv     vs3,32(r5)
    100016d0:	39 00 a5 f5 	lxv     vs45,48(r5)
    100016d4:	70 ff c1 d9 	stfd    f14,-144(r1)
    100016d8:	78 ff e1 d9 	stfd    f15,-136(r1)
    100016dc:	80 ff 01 da 	stfd    f16,-128(r1)
    100016e0:	88 ff 21 da 	stfd    f17,-120(r1)
    100016e4:	90 ff 41 da 	stfd    f18,-112(r1)
    100016e8:	98 ff 61 da 	stfd    f19,-104(r1)
    100016ec:	a0 ff 81 da 	stfd    f20,-96(r1)
    100016f0:	a8 ff a1 da 	stfd    f21,-88(r1)
    100016f4:	b0 ff c1 da 	stfd    f22,-80(r1)
    100016f8:	b8 ff e1 da 	stfd    f23,-72(r1)
    100016fc:	c0 ff 01 db 	stfd    f24,-64(r1)
    10001700:	c8 ff 21 db 	stfd    f25,-56(r1)
    10001704:	d8 51 00 ee 	xvf64ger a4,vs0,vs10
    10001708:	d8 59 80 ee 	xvf64ger a5,vs0,vs11
    1000170c:	d0 ff 41 db 	stfd    f26,-48(r1)
    10001710:	d8 ff 61 db 	stfd    f27,-40(r1)
    10001714:	e0 ff 81 db 	stfd    f28,-32(r1)
    10001718:	e8 ff a1 db 	stfd    f29,-24(r1)
    1000171c:	f0 ff c1 db 	stfd    f30,-16(r1)
    10001720:	f8 ff e1 db 	stfd    f31,-8(r1)
    10001724:	dc 51 80 ed 	xvf64ger a3,vs32,vs10
    10001728:	dc 59 80 ec 	xvf64ger a1,vs32,vs11
    1000172c:	d8 19 00 ef 	xvf64ger a6,vs0,vs3
    10001730:	da 69 80 ef 	xvf64ger a7,vs0,vs45
    10001734:	dc 19 00 ed 	xvf64ger a2,vs32,vs3
    10001738:	de 69 00 ec 	xvf64ger a0,vs32,vs45
    1000173c:	58 00 82 41 	beq     10001794
    10001740:	ff ff 26 39 	addi    r9,r6,-1
    10001744:	a6 03 29 7d 	mtctr   r9
    10001748:	00 00 00 60 	nop
    1000174c:	00 00 00 60 	nop
\end{lstlisting}
    }
    \vspace{-2ex}
    \caption{Object code for the initial part of the DGEMM kernel, including saving of
    nonvolatile registers and initialization of the accumulator through an
    outer-product of the first columns of \code{X} and \code{Y}.}
    \label{Fig:Init}
    \hrule
\end{figure}
}

To conserve space, we limit our discussion to the 
object code for the loop in lines 15--19 of Figure~\ref{Fig:KernelDGEMM},
shown in Figure~\ref{Fig:Loop}. Each column of $X$ is loaded
through two 32-byte load instructions (lines 1--2) and each row of
$Y^T$ is loaded through four 16-byte load instructions (lines
5--8).  The accumulating outer-product of the two 8-element vectors is
implemented by 8 \code{xvf64gerpp} instructions (lines 9--16). Lines
3 and 4 advance the pointers for $Y$ and $X$, respectively.
Line 17 closes the loop.

\begin{figure}[h]
    \hrule
    {\scriptsize
\begin{lstlisting}[language=C,firstline=1182,lastline=1198,escapechar=`,numbers=left,numbersep=0pt]
    10001750:	40 00 a4 19 	lxvp    vs44,64(r4)
    10001754:	60 00 24 18 	lxvp    vs32,96(r4)
    10001758:	40 00 a5 38 	addi    r5,r5,64
    1000175c:	40 00 84 38 	addi    r4,r4,64
    10001760:	09 00 05 f5 	lxv     vs40,0(r5)
    10001764:	19 00 25 f5 	lxv     vs41,16(r5)
    10001768:	29 00 45 f5 	lxv     vs42,32(r5)
    1000176c:	39 00 65 f5 	lxv     vs43,48(r5)
    10001770:	d6 41 0c ee 	xvf64gerpp a4,vs44,vs40
    10001774:	d6 41 80 ed 	xvf64gerpp a3,vs32,vs40
    10001778:	d6 49 8c ee 	xvf64gerpp a5,vs44,vs41
    1000177c:	d6 49 80 ec 	xvf64gerpp a1,vs32,vs41
    10001780:	d6 51 0c ef 	xvf64gerpp a6,vs44,vs42
    10001784:	d6 51 00 ed 	xvf64gerpp a2,vs32,vs42
    10001788:	d6 59 8c ef 	xvf64gerpp a7,vs44,vs43
    1000178c:	d6 59 00 ec 	xvf64gerpp a0,vs32,vs43
    10001790:	c0 ff 00 42 	bdnz    10001750
\end{lstlisting}
    }
    \vspace{-2ex}
    \caption{Object code for the computation loop of the DGEMM kernel. The 
    accumulator is updated by a sequence of $8 \times 8$ outer-products
    of the columns of \code{X} and \code{Y}.}
    \label{Fig:Loop}
    \hrule
\end{figure}

\commentout{
Object code for the final part of the DGEMM kernel (lines 21--28 of Figure~\ref{Fig:KernelDGEMM})
are shown in Figure~\ref{Fig:Save}. The values of the accumulators are first
transferred to the corresponding vector-scalar registers (lines 1--8) and then
stored to memory (lines 9--40). Finally, the nonvolatile registers are restored
(lines 41--58) and the function returns (line 59).

\begin{figure}[h]
    \hrule
    {\scriptsize
\begin{lstlisting}[language=C,firstline=1199,lastline=1257,escapechar=`,numbers=left,numbersep=0pt]
    10001794:	62 01 00 7e 	xxmfacc a4
    10001798:	62 01 80 7e 	xxmfacc a5
    1000179c:	62 01 00 7f 	xxmfacc a6
    100017a0:	62 01 80 7f 	xxmfacc a7
    100017a4:	62 01 80 7d 	xxmfacc a3
    100017a8:	62 01 80 7c 	xxmfacc a1
    100017ac:	62 01 00 7d 	xxmfacc a2
    100017b0:	62 01 00 7c 	xxmfacc a0
    100017b4:	05 00 03 f6 	stxv    vs16,0(r3)
    100017b8:	15 00 23 f6 	stxv    vs17,16(r3)
    100017bc:	25 00 43 f6 	stxv    vs18,32(r3)
    100017c0:	35 00 63 f6 	stxv    vs19,48(r3)
    100017c4:	45 00 83 f6 	stxv    vs20,64(r3)
    100017c8:	55 00 a3 f6 	stxv    vs21,80(r3)
    100017cc:	65 00 c3 f6 	stxv    vs22,96(r3)
    100017d0:	75 00 e3 f6 	stxv    vs23,112(r3)
    100017d4:	85 00 03 f7 	stxv    vs24,128(r3)
    100017d8:	95 00 23 f7 	stxv    vs25,144(r3)
    100017dc:	a5 00 43 f7 	stxv    vs26,160(r3)
    100017e0:	b5 00 63 f7 	stxv    vs27,176(r3)
    100017e4:	c5 00 83 f7 	stxv    vs28,192(r3)
    100017e8:	d5 00 a3 f7 	stxv    vs29,208(r3)
    100017ec:	e5 00 c3 f7 	stxv    vs30,224(r3)
    100017f0:	f5 00 e3 f7 	stxv    vs31,240(r3)
    100017f4:	35 01 e3 f5 	stxv    vs15,304(r3)
    100017f8:	05 01 83 f5 	stxv    vs12,256(r3)
    100017fc:	15 01 a3 f5 	stxv    vs13,272(r3)
    10001800:	25 01 c3 f5 	stxv    vs14,288(r3)
    10001804:	45 01 83 f4 	stxv    vs4,320(r3)
    10001808:	55 01 a3 f4 	stxv    vs5,336(r3)
    1000180c:	65 01 c3 f4 	stxv    vs6,352(r3)
    10001810:	75 01 e3 f4 	stxv    vs7,368(r3)
    10001814:	85 01 03 f5 	stxv    vs8,384(r3)
    10001818:	95 01 23 f5 	stxv    vs9,400(r3)
    1000181c:	a5 01 43 f5 	stxv    vs10,416(r3)
    10001820:	b5 01 63 f5 	stxv    vs11,432(r3)
    10001824:	c5 01 03 f4 	stxv    vs0,448(r3)
    10001828:	d5 01 23 f4 	stxv    vs1,464(r3)
    1000182c:	e5 01 43 f4 	stxv    vs2,480(r3)
    10001830:	f5 01 63 f4 	stxv    vs3,496(r3)
    10001834:	70 ff c1 c9 	lfd     f14,-144(r1)
    10001838:	78 ff e1 c9 	lfd     f15,-136(r1)
    1000183c:	80 ff 01 ca 	lfd     f16,-128(r1)
    10001840:	88 ff 21 ca 	lfd     f17,-120(r1)
    10001844:	90 ff 41 ca 	lfd     f18,-112(r1)
    10001848:	98 ff 61 ca 	lfd     f19,-104(r1)
    1000184c:	a0 ff 81 ca 	lfd     f20,-96(r1)
    10001850:	a8 ff a1 ca 	lfd     f21,-88(r1)
    10001854:	b0 ff c1 ca 	lfd     f22,-80(r1)
    10001858:	b8 ff e1 ca 	lfd     f23,-72(r1)
    1000185c:	c0 ff 01 cb 	lfd     f24,-64(r1)
    10001860:	c8 ff 21 cb 	lfd     f25,-56(r1)
    10001864:	d0 ff 41 cb 	lfd     f26,-48(r1)
    10001868:	d8 ff 61 cb 	lfd     f27,-40(r1)
    1000186c:	e0 ff 81 cb 	lfd     f28,-32(r1)
    10001870:	e8 ff a1 cb 	lfd     f29,-24(r1)
    10001874:	f0 ff c1 cb 	lfd     f30,-16(r1)
    10001878:	f8 ff e1 cb 	lfd     f31,-8(r1)
    1000187c:	20 00 80 4e 	blr
\end{lstlisting}
    }
    \vspace{-2ex}
    \caption{Object code for the final part of the DGEMM kernel, 
    including the transfer of accumulator values, the store of those values
    and the restoring of the nonvolatile registers.}
    \label{Fig:Save}
    \hrule
\end{figure}
}

\subsection{{\sf SCONV}}
\label{Sec:SCONV}

The characteristics of a convolution operation are described by a
variety of parameters, including size of kernel, amount of padding, step
increments, etc. In this section, we consider a simple two-dimensional
convolution to illustrate this kind of computation using the new MMA
instructions.

Let $h$ be a $3 \times 3$ kernel, and $A$ an $m \times n$ image,
both represented as matrices:
\begin{eqnarray}
    h 	& =	&
    \left[
	\begin{array}{ccc}
	    h_{0}	& h_{1}	& h_{2}	\\
	    h_{3}	& h_{4}	& h_{5}	\\
	    h_{6}	& h_{7}	& h_{8}
	\end{array}
    \right],
    \\ 
    A 	& =	&
    \left[
	\begin{array}{cccc}
	    a_{0,0}	& a_{0,1}	& \cdots	& a_{0,n-1}	\\
	    a_{1,0}	& a_{1,1}	& \cdots	& a_{1,n-1}	\\
	    \vdots	& \vdots	& \ddots	& \vdots	\\
	    a_{m-1,0}	& a_{m-1,1}	& \cdots	& a_{m-1,n-1}	
	\end{array}
    \right].
\end{eqnarray}
We want to compute the $(m-2) \times (n-2)$ matrix $C = h * A$ which is
the convolution of kernel $h$ with the image $A$ (no padding, single
stepping in both dimensions). Let $C_i, i=0,\ldots,m-3$ denote the
$i$-th row of matrix $C$, which can be expressed as a vector-matrix
multiplication:

\begin{eqnarray}
    C_i	=
    \left[
	\begin{array}{ccccccccc}
	    h_{0}	& h_{1}	& h_{2}	& h_{3}	& h_{4}	& h_{5}	& h_{6}	& h_{7}	& h_{8}
	\end{array}
    \right]
	\times
    \bar{A}_i 
\end{eqnarray}
where $\bar{A}_i$ is a $9 \times (m-2)$ matrix derived from $A$:
\begin{eqnarray}
    \bar{A}_i	& = 	&
    \left[
	\begin{array}{cccc}
	    a_{i+0,0}	& a_{i+0,1}	& \cdots	& a_{i+0,n-3}	\\
	    a_{i+0,1}	& a_{i+0,2}	& \cdots	& a_{i+0,n-2}	\\
	    a_{i+0,2}	& a_{i+0,3}	& \cdots	& a_{i+0,n-1}	\\
	    a_{i+1,0}	& a_{i+1,1}	& \cdots	& a_{i+1,n-3}	\\
	    a_{i+1,1}	& a_{i+1,2}	& \cdots	& a_{i+1,n-2}	\\
	    a_{i+1,2}	& a_{i+1,3}	& \cdots	& a_{i+1,n-1}	\\
	    a_{i+2,0}	& a_{i+2,1}	& \cdots	& a_{i+2,n-3}	\\
	    a_{i+2,1}	& a_{i+2,2}	& \cdots	& a_{i+2,n-2}	\\
	    a_{i+2,2}	& a_{i+2,3}	& \cdots	& a_{i+2,n-1}	
	\end{array}
    \right].
    \label{Eq:Conv3x3}
\end{eqnarray}
Matrix $A_i$ is formed by three rows of $A$, each appearing three times:
once in its original form, once shifted left by one position and once
shifted left by two positions.

It is common to apply multiple convolution kernels to the same input
image. This can be accomplished in parallel by building a matrix $\bar{H}$
with the parameters of each kernel as a row of the matrix. If there are
$k$ kernels, then $\bar{H}$ is a $k \times 9$ matrix that multiples a
$9 \times (m-2)$ matrix and we have transformed our convolution into a
(series of) matrix multiplication(s).

Quite often, an image has multiple channels (\eg, $R$, $G$, and $B$
channels for the red, green and blue components, respectively).  The
multiple channels can be concatenated to form a ``taller'' $\bar{A}_i$
matrix, which is then multiplied by a ''wider'' $\bar{H}$ matrix. In the
3-channel case, We end up with a $k \times 27$ by $27 \times (m-2)$ matrix
multiplication. This produces $k$ rows of output, one for each kernel.

When using an existing matrix-multiplication service, either a hardware
engine directly or a {\sf GEMM} routine in a BLAS library, one has to
materialize the $\bar{A}_i$ matrices so that matrix multiplication
can be invoked. (Some deep learning accelerators, like Google's
TPU~\cite{10.1145/3079856.3080246}, avoid this overhead by supporting
convolution directly in hardware. The additional hardware required is not
that significant, but it goes beyond just a matrix multiplication
engine.)

With the fine-grain instructions in the MMA facility, convolution can be
done directly on the input matrix $A$ by using code that is similar to
the {\sf DGEMM} code discussed in Section~\ref{Sec:DGEMM}.  The $\bar{H}$
matrix plays the role of the left matrix and can be prepared in advance,
since a kernel is typically applied repeatedly to incoming data. The
image matrix $A$ plays the role of the right matrix, but each of its rows
is loaded three times, each time starting at a different displacement.
Once a column of $\bar{H}$ and a row of $A$ are loaded in the processors,
their outer product can be computed with the same MMA instructions that
would be used for a matrix multiplication.

The code for a 3-channel $3 \times 3$ convolution kernel is shown
in Figures~\ref{Fig:MacrosSCONV} and \ref{Fig:KernelSCONV}.
Figure~\ref{Fig:MacrosSCONV} shows the supporting definitions,
similar to the ones for {\sf DGEMM}. The data type in this
case is 32-bit single-precision floating point. The 8
architected accumulators, each holding a $4 \times 4$ array
of \code{float}s, are used to form an $8 \times 16$
virtual accumulator. Each update operation consists 
of an $8 \times 16$ outer product that is added to the accumulator.

\begin{figure}
    \hrule
    {\scriptsize
\begin{lstlisting}[language=C,firstline=11,lastline=38,escapechar=`,numbers=left,numbersep=2pt]
typedef	__vector_quad		fp32_4x4;
typedef __vector unsigned char	fp32_4;

#define mma_store_acc(AS, A, D)			\
{						\
    fp32_4	a[4];				\
    __builtin_mma_disassemble_acc(a,&(AS));	\
    *((fp32_4*)A+D+0) = a[0];			\
    *((fp32_4*)A+D+1) = a[1];			\
    *((fp32_4*)A+D+2) = a[2];			\
    *((fp32_4*)A+D+3) = a[3];			\
}

#define mma_xvf32_8x16(acc, op, X, Y)			\
{							\
    fp32_4	x0, x1;					\
    fp32_4	y0, y1, y2, y3;				\
    x0 = *((fp32_4*)X+0); x1 = *((fp32_4*)X+1);		\
    y0 = *((fp32_4*)Y+0); y1 = *((fp32_4*)Y+1); 	\
    y2 = *((fp32_4*)Y+2); y3 = *((fp32_4*)Y+3); 	\
    __builtin_mma_xvf32 ## op (&(acc[0]), x0, y0);	\
    __builtin_mma_xvf32 ## op (&(acc[1]), x0, y1);	\
    __builtin_mma_xvf32 ## op (&(acc[4]), x1, y0);	\
    __builtin_mma_xvf32 ## op (&(acc[5]), x1, y1);	\
    __builtin_mma_xvf32 ## op (&(acc[2]), x0, y2);	\
    __builtin_mma_xvf32 ## op (&(acc[3]), x0, y3);	\
    __builtin_mma_xvf32 ## op (&(acc[6]), x1, y2);	\
    __builtin_mma_xvf32 ## op (&(acc[7]), x1, y3);	\
\end{lstlisting}
    }
    \vspace{-2ex}
    \caption{Supporting definitions for the {\sf SCONV} kernel code.}
    \label{Fig:MacrosSCONV}
    \hrule
\end{figure}

The convolution kernel itself is shown in Figure~\ref{Fig:KernelSCONV}.
The $R$, $G$ and $B$ matrices are the the input channels, and $H$ is
the matrix of kernels. $C$ is the output matrix and $n$ the number of
elements per row of the input matrices. There are a total of 27 outer
product operations.  Three rows from each channel are used three times
each, as shown in Equation~\ref{Eq:Conv3x3}.

\begin{figure}
    \hrule
    {\scriptsize
\begin{lstlisting}[language=C,firstline=41,lastline=103,escapechar=`,numbers=left,numbersep=2pt]
void sconv_kernel_8x27x16
(
    float		*C,
    const float		*H,
    const float 	*R,
    const float 	*G,
    const float 	*B,
    const uint64_t	 n
)
{
    fp32_4x4	acc[8];

    mma_xvf32_8x16(acc, ger  , H+  0, R+0);
    mma_xvf32_8x16(acc, gerpp, H+  8, R+1);
    mma_xvf32_8x16(acc, gerpp, H+ 16, R+2);

    R += n;
    mma_xvf32_8x16(acc, gerpp, H+ 24, R+0);
    mma_xvf32_8x16(acc, gerpp, H+ 32, R+1);
    mma_xvf32_8x16(acc, gerpp, H+ 40, R+2);

    R += n;
    mma_xvf32_8x16(acc, gerpp, H+ 48, R+0);
    mma_xvf32_8x16(acc, gerpp, H+ 56, R+1);
    mma_xvf32_8x16(acc, gerpp, H+ 64, R+2);

    mma_xvf32_8x16(acc, gerpp, H+ 72, G+0);
    mma_xvf32_8x16(acc, gerpp, H+ 80, G+1);
    mma_xvf32_8x16(acc, gerpp, H+ 88, G+2);

    G += n;
    mma_xvf32_8x16(acc, gerpp, H+ 96, G+0);
    mma_xvf32_8x16(acc, gerpp, H+104, G+1);
    mma_xvf32_8x16(acc, gerpp, H+112, G+2);

    G += n;
    mma_xvf32_8x16(acc, gerpp, H+120, G+0);
    mma_xvf32_8x16(acc, gerpp, H+128, G+1);
    mma_xvf32_8x16(acc, gerpp, H+136, G+2);

    mma_xvf32_8x16(acc, gerpp, H+144, B+0);
    mma_xvf32_8x16(acc, gerpp, H+152, B+1);
    mma_xvf32_8x16(acc, gerpp, H+160, B+2);

    B += n;
    mma_xvf32_8x16(acc, gerpp, H+168, B+0);
    mma_xvf32_8x16(acc, gerpp, H+176, B+1);
    mma_xvf32_8x16(acc, gerpp, H+184, B+2);

    B += n;
    mma_xvf32_8x16(acc, gerpp, H+192, B+0);
    mma_xvf32_8x16(acc, gerpp, H+200, B+1);
    mma_xvf32_8x16(acc, gerpp, H+208, B+2);

    mma_store_acc(acc[0], C, 0);
    mma_store_acc(acc[1], C, 4);
    mma_store_acc(acc[2], C, 8);
    mma_store_acc(acc[3], C,12);
    mma_store_acc(acc[4], C,16);
    mma_store_acc(acc[5], C,20);
    mma_store_acc(acc[6], C,24);
    mma_store_acc(acc[7], C,28);
}
\end{lstlisting}
    }
    \vspace{-2ex}
    \caption{{\sf SCONV} $8 \times 27 \times 16$ kernel code.}
    \label{Fig:KernelSCONV}
    \hrule
\end{figure}

\section{Performance measurements}
\label{Sec:Performance}

We evaluate the performance of a POWER10 core with the matrix math
engine using the University of Tennessee High Performance Linpack (HPL)
benchmark~\cite{HPL}.  HPL is a computation intensive benchmark with most (over 90\%
for large enough problems) of execution time spent on a double-precision
matrix multiply kernel (DGEMM) and much of the rest in other BLAS kernels.
We use the standard OpenBLAS in our distribution of Linux but we hand
write the DGEMM kernel for the critical size $M = 128$, $N = 128$, $K =
128$. The code is based on the example of Figure~\ref{Fig:KernelDGEMM},
with aggressive unrolling and other optimizations to reduce overhead.

Our experiments are performed on a preliminary hardware POWER10
platform, running at reduced clock frequency. (Both the nest and
the core run at this reduced frequency.) We also use a commercially
available POWER9 system and perform three types of measurements of
single-core, single-thread performance: (1) We run a POWER9-compliant
code that only uses POWER9 ISA instructions (vector instructions) on
the commercially available POWER9 system. (2) We run exactly the same
code on our preliminary POWER10 platform (labeled {\sf POWER10-VSX}. (3)
We run the MMA-enabled code on our preliminary POWER10 platform (labeled
{\sf POWER10-MMA}).

\begin{figure}[htb]
    \hrule
    \vspace{-2ex}
    \includegraphics[trim=50 0 50 0, clip, width=0.50\textwidth]{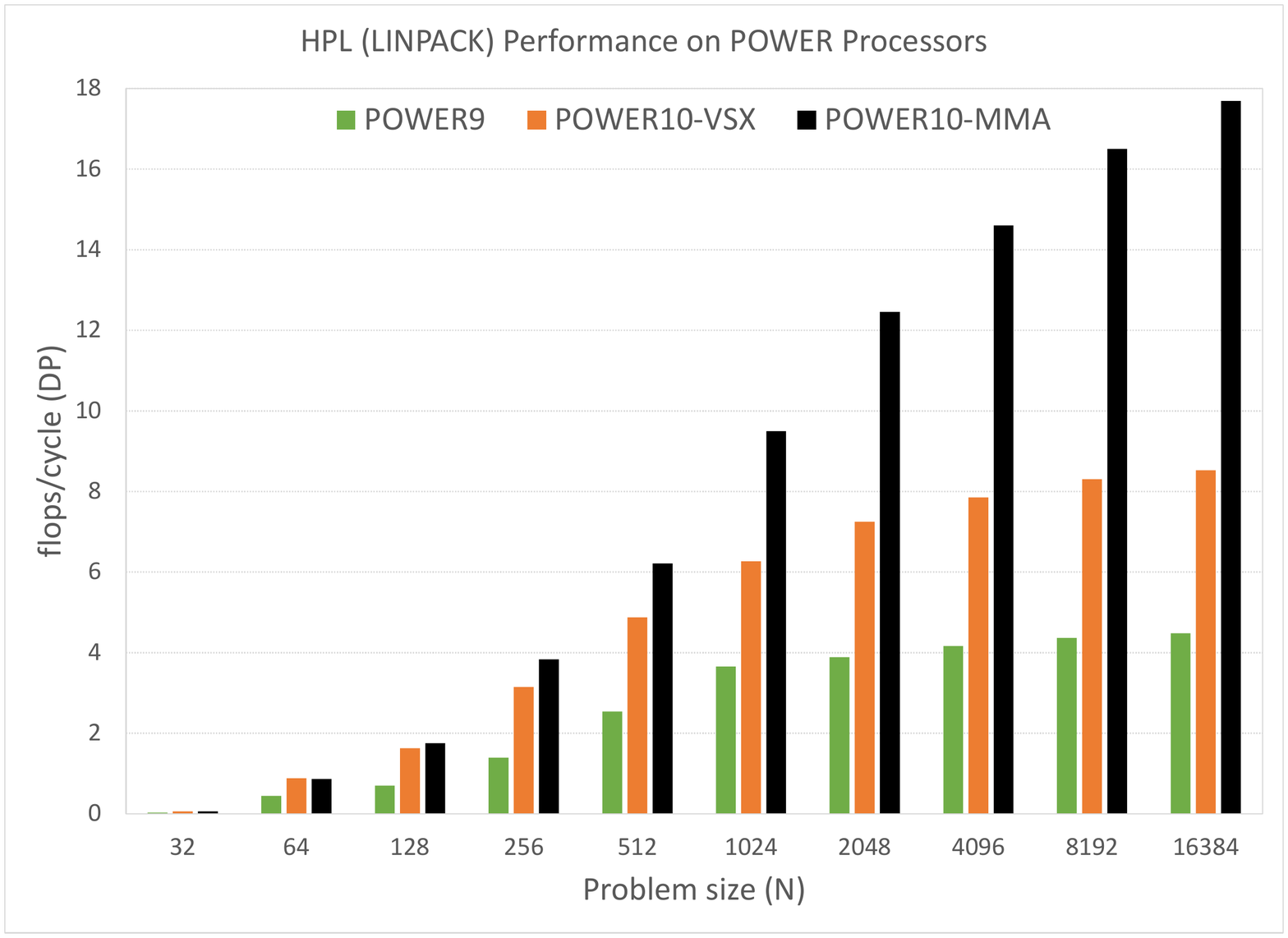}
    \vspace{-7ex}
    \caption{HPL (Linpack) performance on POWER9 and POWER10 processors.}
    \label{Fig:HPL}
    \hrule
\end{figure}

Results for the three cases are shown in Figure~\ref{Fig:HPL} as
a function of problem size. Performance is reported in flops/cycle.
As expected, overall performance increases with problem size, as a higher
percentage of the computation is contained within the key $128 \times 128$
DGEMM.  For the larger problem sizes, vector code in POWER10 outperforms
the same vector code in POWER9 by a factor of two.  The POWER10 advantage
is explained by it having four vector pipelines as opposed to only two in
POWER9.  The MMA code in POWER10 outperforms the vector code in the same
platform also by a factor of two \hl{($4 \times$ better than POWER9)}.
This is inline with the throughput of the two matrix pipelines being
double that of the four vector pipes.  \hl{For more performance results
on both HPL and ResNet-50 (also $4 \times$ the per core performance of
POWER9), we refer the reader to}~\cite{9352481}\hl{.}

\begin{figure}[htb]
    \hrule 
    \vspace{-2ex} 
    \includegraphics[trim=50 0 50 0, clip, width=0.50\textwidth]{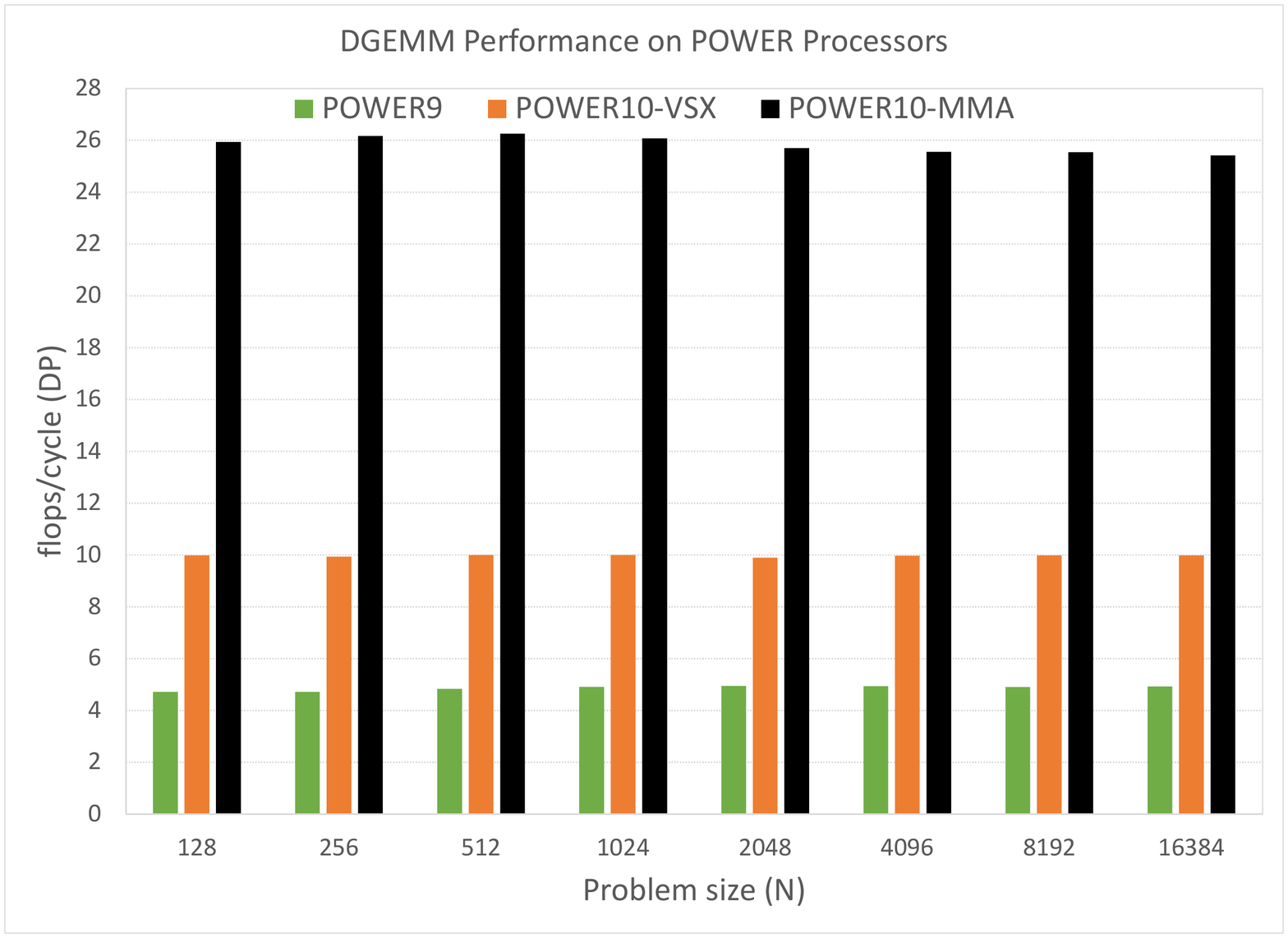}
    \vspace{-7ex} 
    \caption{\hl{DGEMM performance on POWER9 and POWER10 processors, 
    multiplying $N \times 128$ matrix $A$ and $128 \times N$ matrix $B$.}} 
    \label{Fig:DGEMM} 
    \hrule
\end{figure}

We take a closer look at the performance of the $128 \times 128$ DGEMM
kernel in Figure~\ref{Fig:DGEMM}. We measure the performance of an $N
\times 128$ by $128 \times N$ matrix multiplication for various values
of N. The result is an $N \times N$ matrix and the computation makes
extensive use of our $128 \times 128$ {\sf DGEMM} kernel.

The vector code running on the POWER9 platform achieves approximately 4.5
flops/cycle, which is 56\% of the peak of 8 flops/cycle in that system.
The same vector code achieves almost 10 flops/cycle (or 62\% of the vector
peak) on the POWER10 platform. Finally, the matrix math engine code
achieves close to 26 flops/cycle (over 80\% of peak) on POWER10.  The
improved efficiency comes from the more natural fit of the instructions
to the computation, as discussed in Section~\ref{Sec:POWER10}. Combined
with the increased throughput of the matrix math engine, we achieve more
than 2.5 times the performance of the vector code on POWER10 and more
than 5.5 times the performance of the vector code on POWER9. \hl{These
are better than the $4 \times$ gains in HPL}, since the rest of that
benchmark code does not leverage the matrix math engine.

\section{Power efficiency aspects}
\label{Sec:Power}

We evaluate the power efficiency of our
approach using a simulation-based IBM internal power
methodology~\cite{DBLP:conf/iccad/DhanwadaHZPMDJRG13}. We run the same
code used for performance evaluation (Section~\ref{Sec:Performance}), in
single-thread mode, through a detailed pre-silicon model of the core. We
capture multiple 5000-instruction windows and evaluate the power draw
during each window.  We then average across all the windows. We measure
power draw for the core without the matrix math engine, as well as just
the matrix math engine itself.

\begin{figure}[htb]
    \hrule \vspace{-2ex} 
    \includegraphics[trim=50 0 50 0, clip, width=0.50\textwidth]{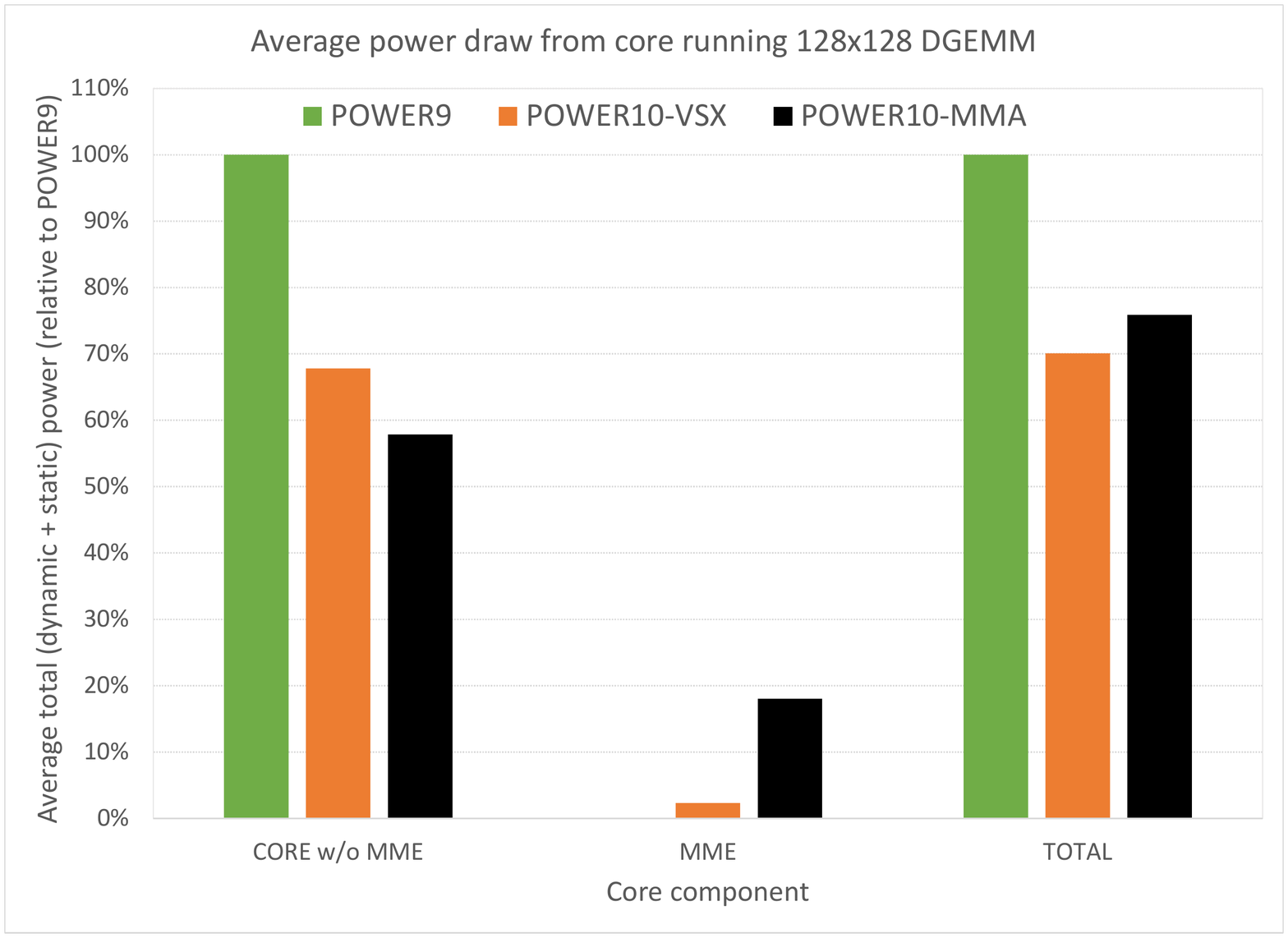}
    \vspace{-7ex} 
    \caption{Average power draw of $128 \times 128$ DGEMM on POWER9 and POWER10 processors.} 
    \label{Fig:power} 
    \hrule
\end{figure}

The average power draw of a POWER9 and POWER10 core during
execution of a $128 \times 128$ {\sf DGEMM} computation are shown in
Figure~\ref{Fig:power}.  For each configuration  (POWER9, POWER10 with
VSX code, POWER10 with MMA code, all running at the same frequency)
we show the average power of a processor core without the matrix math
engine ({\sf CORE w/o MME}), just the matrix math engine ({\sf MME}) and total
({\sf TOTAL} -- the sum of the two).  The POWER9 core does not have an
\mme, so the total is the same as just the core.

Comparing Figures~\ref{Fig:DGEMM} and \ref{Fig:power} we observe that
the POWER10 core running MMA code delivers $2.5 \times$ the performance
of the same core running VSX code, while drawing only 8\% more power.
When the MME unit is power gated, thus eliminating any \mme\ draw when
running the VSX code, that difference increases to 12\%. In any case,
it is a small power increase for a significant boost in performance.
If we compare to the previous generation POWER9 core, which uses an older
silicon technology, we achieve a $5 \times$ improvement in performance
at 24\% less power. This corresponds to almost $7 \times$ reduction on
energy per computation, looking at the core level.

\section{Conclusions}
\label{Sec:Conclusions}

The MMA facility is a new addition to the Power ISA\texttrademark\
Version 3.1 that will appear in future IBM POWER processors. The
facility adds a set of instructions tailored for matrix math, directly
implementing rank-$k$ update of small matrices of 32-bit signed integers
(with mixed-precision inputs), single-precision floating-point and
double-precision floating-point numbers.  The new MMA instructions are
a significant departure from current vector instruction sets, which
typically operate on homogeneous vector registers. The MMA instructions
use vector registers as inputs, but update a new set of registers called
\emph{accumulators}.

It will take time for compilers to catch up with automatic code generation
for the MMA facility. Meanwhile, we have augmented the GNU Compiler
Collection, and are in process of augmenting LLVM-based compilers,
with a new set of \builtins\ that match the functionality of the MMA
facility. These \builtins\ give the programmers great control of the
generated code while freeing them from details of register allocation
and instruction scheduling.  The source code using \builtins\ is easier
to write and maintain than assembly code, and the generated object code
is efficient, with few or no additional overhead instructions.

The matrix math engine in the IBM POWER10 processor core is the first to
implement the new MMA facility instructions. It has fully achieved its
objective of quadrupling the computation rate of matrix multiplication
kernels over its POWER9 predecessor.  That improvement has translated
well to code that makes heavy use of those kernels, such as HPL.

The fine-grain nature of the MMA facility instructions mean that they can
be used for various computations. We have shown in this paper how they
fit into matrix multiplication and convolution. Other research work is
exploring their use in stencil computations and discrete Fourier transform.

Code leveraging the MMA instructions is already included in
OpenBLAS \hl{and Eigen}, and can be built using the most recent versions of GCC.
The uploaded OpenBLAS code supports double, single and half (\code{bf16})
precision floating-point. The new MMA instructions are present in the
matrix-multiply ({\sf GEMM}) kernels, which are used as building blocks
for various BLAS routines.

Much of our future work consists of extending the applicability of the
new MMA instructions. In addition to compiler support for automatic MMA
code generation, we are also investigating what architectural features
can make the MMA useful for a broader set of applications.

\section*{Acknowledgements}
We want to thank all our colleagues who worked on the research
and development of the new IBM POWER10 system. This work would
not be possible without their extreme dedication and effort.

\pagebreak

\bibliographystyle{IEEEtranS}
\bibliography{main}

\commentout
{
\appendix
\section{Artifact reproduction}

The compiler used in this study can be obtained by downloading
and installing IBM Advance Toolchain version 15.0 (alpha) 
with the following commands (the default install location is \code{/opt}):
\begin{verbatim}
git clone https://github.com/advancetoolchain/advance-toolchain.git
cd advance-toolchain
make AT_CONFIGSET=next BUILD_ARCH=ppc64le
\end{verbatim}
The tool chain can be installed either as a native
tool chain, running on Linux on Power systems, or as
a cross-compilation tool chain on Linux on x86 systems.

POWER10 executables can be run on the IBM POWER10 Functional Level Simulator, 
which includes support for all MMA
instructions. This simulator runs the object code 
produced by the above compiler.
It can be downloaded from
\begin{verbatim}
https://www14.software.ibm.com/webapp/set2/sas/f/pwrfs/pwr10/home.html
\end{verbatim}
}

\end{document}